\documentclass{emulateapj}
\newcommand{\ee}[1]{\mbox{${} \times 10^{#1}$}}
\newcommand{\msun}{\mbox{$M_\odot$}}

\def \numberofruns      {\the\count100\ }
\def \numberofrunstotal {\the\count101\ }

\begin{document}

\title{The Fate of Dwarf Galaxies in Clusters and the Origin of 
Intracluster Stars. I. Isolated Clusters}

\author{Paramita Barai,\altaffilmark{1}
William Brito,\altaffilmark{1} and
Hugo Martel\altaffilmark{1}}

\altaffiltext{1}{D\'epartement de physique, de g\'enie physique et d'optique,
Universit\'e Laval, Qu\'ebec, QC, Canada}

\begin{abstract}

The main goal of this paper is to compare the relative
importance of destruction by tides, vs. destruction
by mergers, in order to assess if tidal destruction of dwarf galaxies 
in clusters is a viable scenario for explaining the origin of intracluster 
stars.
We have designed a simple algorithm for simulating the evolution of
isolated clusters. The distribution of galaxies in the cluster is evolved
using a direct gravitational N-body algorithm combined with a subgrid
treatment of physical processes such as
mergers, tidal disruption, and galaxy harassment. Using this
algorithm, we have performed a total of \numberofrunstotal simulations.
Our main results are (1) destruction of dwarf galaxies by mergers
dominates over destruction by tides, and (2) the destruction of
dwarf galaxies by tides is sufficient to explain the observed
intracluster light in clusters.

\end{abstract}


\section{INTRODUCTION} 
\label{sec-intro} 

\subsection{Dwarf Galaxies}

Dwarf galaxies (DGs) are the most numerous galaxies occurring in the Universe. 
A majority of galaxies in the local group are DGs \citep{mateo98}. 
Also DGs comprise 85\% of the Local Volume galaxy population 
\citep[$D \leq 10$ Mpc,][]{karachentsev04}, and
have been seen in observations of nearby galaxy clusters, 
Coma \citep{thompson93, bernstein95}, 
Virgo \citep{sandage85, impey88, phillipps98, lee2003}, 
Fornax \citep{bothun91, drinkwater03}, Centaurus \citep{mieske07}, 
and several galaxy groups 
\citep{karachentseva85, ferguson91, cote97, carrasco01, cellone05}. 
DGs may have a space density $\sim 40$ times that of bright galaxies in the 
Universe \citep{staveley92}. 

DGs are defined as low-mass ($10^7 - 10^9$ \msun) galaxies 
having an absolute magnitude fainter than 
$M_B \sim -16$ mag, or $M_V \sim -18$ mag \citep{grebel01}, 
have low surface brightness and low metallicity. 
Their small stellar fraction and very low luminosities
make them the hardest galaxies to detect. 
They are believed to be the single systems with the largest 
proportion of dark-matter, 
and have a correspondingly
high ratio of dark to luminous mass \citep[e.g.,][]{cote00}, 
with $M/L$ ratios as high as that of galaxy groups and poor clusters. 
Due to their smaller masses and gravitational potentials, 
DGs are less able to retain their gas as compared to more massive galaxies, and 
in a clustered environment,
the DGs are more likely to be disrupted by galactic encounters 
and environmental effects. 

In the hierarchical clustering scenario of structure formation in the Universe
\citep[e.g.,][]{white91, kauffmann93}, the accretion of DGs causes
the build up and growth of massive galaxies and large-scale structures. 
There exists a deficiency in the number of observed low-luminosity DGs 
(discrepancy more than one order of magnitude) as compared to the 
large number of theoretically predicted low-mass dark matter halos 
\citep[e.g.,][]{trentham02b, trentham06}. 
This is recognized as a problem for cold dark matter theory,
and the likely solution to this problem involves energy
feedback from stellar evolution.
Dense clusters are observed to contain a larger number of low-luminosity DGs 
per high-luminosity giant galaxy when compared to the field. 
\citet{trentham02a} found that the Virgo cluster contains $\sim 2.5$ 
times more 
dwarf galaxies per giant galaxy when compared to the Ursa Major cluster. 
These imply that dwarfs are more common relative to giants 
in dense environments than diffuse ones. 

The dynamical evolution of galaxies in a cluster is influenced 
by several mechanisms. 
There are two types of tidal interactions: 
the tidal forces due to other (massive) cluster galaxies \citep{gnedin03}, 
and the tidal field resulting from the overall cluster potential 
\citep{merritt84, byrd90}. 
There can be collisions between the galaxies themselves due to their motion,
sometimes resulting in mergers. 
Of particular interest is the occurrence of multiple high-velocity encounters 
between cluster galaxies \citep{richstone76}, 
a phenomenon which is termed ``galaxy harassment'' \citep{moore96}. 
A phenomenon like tidal stirring \citep[][where tidal shocks strip DGs]{mayer01} 
has a more pronounced effect on the less massive galaxies. 
There can also be stripping within galaxy groups and protoclusters 
accreting onto a cluster \citep{mihos04}. 

Works are found in the literature on the interaction of cluster 
environment with DGs in the cluster. 
Studying the core of the Fornax cluster, \citet{hilker99} 
examined a scenario in which DGs are accreted and dissolved at the 
cluster center. 
They found that the infall of DGs can largely explain many properties, 
but there are probably other physical processes occurring simultaneously, 
like the evolution of the more massive galaxies by stripping and merging. 
\citet{mori00} performed analytical and hydrodynamic studies of the 
interaction between the ICM of a cluster and an extended gas component of DGs 
surrounded by a cold dark matter halo. 
They found that the DG halos lose their diffuse gas rapidly by 
ram-pressure stripping in a typical cluster environment. 

\subsection{Intracluster Light}

The diffuse intracluster light (ICL) observed in clusters of galaxies
is produced by stars, usually of low surface brightness, located
outside individual galaxies but within the cluster
and associated with the cluster potential. 
The first mention of IC light was made by \citet{zwicky51}, 
who observed extended irregular regions of stars and low surface-brightness matter 
in the intergalactic spaces of the Coma cluster. 
Several observations have since detected diffuse light in 
galaxy clusters 
\citep[e.g.,][for reviews]{vilchez99, arnaboldi04}. 
Diffuse ICL has been observed in many galaxy cluster systems 
\citep[e.g.,][]{gonzalez05, krick06}, 
including non-cD galaxy clusters \citep[e.g.,][]{feldmeier2004a}. 
Observations of diffuse ICL in Virgo \citep{mihos05} reveal complex 
substructures in
intricate web patterns including long tidal streamers, small tidal 
tails, and intergalactic bridges. 
The idea of IC globular clusters was proposed by \citet{west95}. 
Later on, distinct IC stars were observed, including globular clusters, 
red giant stars, and SN Ia \citep{galyam03}. 
Individual IC stars have been found in 
Virgo \citep{ferguson98, arnaboldi03}, and Coma \citep{gerhard05}. 
Deep broadband imaging has helped to observe ICL in several other 
local galaxy clusters, 
e.g., Cantaurus, Fornax, Abell, and clusters up to intermediate redshifts. 
The origin and evolution of the IC stars and diffuse light are not well 
constrained at present. 

Arguably these field stars are gravitationally stripped from their parent 
galaxies when the galaxies in a cluster interact dynamically with other galaxies 
or with the cluster potential. 
Continuous accumulation of ICL is now believed to be 
an ubiquitous feature of evolved galaxy clusters, 
with the unbound (i.e. intracluster) star fraction slowly increasing 
with time. {From} observations and cosmological simulations, 
at $z=0$ at least $10 - 20 \%$ of all stars in a cluster are 
unbound to any one galaxy \citep[e.g.,][]{aguerri05}. 
As an upper limit, 
ICL constitute $\approx 10 - 40 \%$ of a cluster's luminosity.
In observations and simulations, 
the fraction of stars in ICL increases with mass of the clusters,
and increases with density of environment: 
from loose groups ($< 2 \%$, \citealt{castro-rodriguez03}), 
to Virgo-like (10\%, \citealt{feldmeier2004b,zibetti05}) and 
rich clusters ($\sim20\%$ or higher, 
\citealt{tyson95,feldmeier2004a,krick07}). 
In the cores of dense and rich clusters (like Coma) 
the local ICL fraction can be as high as $50 \%$ \citep{bernstein95}. 

The most popular formation mechanism of IC population is stripping 
of stars from cluster galaxies by gravitational tides, fast
encounters between galaxies, 
and tidal interactions between colliding and merging galaxies 
\citep{miller83, richstone83, weil97, gregg98}. 
Another contribution to the IC population can come from the 
infall of galaxy protoclusters containing stars which are already unbound. 
Studies of IC stars in the Virgo cluster \citep{feldmeier2004b} 
imply that bulk of the IC stars come from late-type galaxies 
and dwarf galaxies, 
and that the IC stars form by tidal stripping. 
Optical studies of ICL in intermediate redshift clusters \citep{zibetti05} 
support that ICL is produced by stripping and disruption of galaxies 
when they pass through cluster centers. 
A dominant mechanism of ICL formation is believed to be 
tidal stripping during the hierarchical assembly of clusters. 

Several authors have studied the origin of the diffuse ICL in clusters
of galaxies using numerical simulation.
\citet{napolitano03} performed high-resolution N-body $\Lambda$CDM simulations 
of a Virgo-like cluster to study the velocity and clustering properties of the 
IC stellar component at $z = 0$. 
The member galaxies in their simulated cluster undergo tidal interactions among 
themselves and with the cluster potential, to produce diffuse stars in the ICM . 
Their results substantially agree 
with the observed clustering properties of the diffuse IC stars in Virgo. 
Using hydrodynamical simulation of a (192 $h^{-1}$ Mpc)$^3$ cosmological box, 
\citet{murante04} studied the statistical properties of IC stars in galaxy clusters 
and the dependence of the ICL properties on cluster mass and temperature. 
Their simulations reveal substantial (at least $\sim 10\%$) 
diffuse stellar light in a cluster, 
hidden from observations because of its very low surface brightness. 
Using high-resolution N-body + SPH simulations of a Coma-like cluster 
formed in a cosmological context, \citet{willman04} studied 
the formation, evolution and properties of IC stars in a rich cluster. 
Their results indicate that both massive and smaller galaxies 
contribute to ICL formation, with the stars stripped preferentially 
from the outer, lower-metallicity parts of a galaxy's stellar distribution. 
The resulting fraction ($\sim 20 \%$ at $z= 0$), and distribution of IC stars 
are in good agreement with the observed ICL properties. 
\citet{sommerlarsen05} performed cosmological TREESPH simulations 
of the formation and evolution of galaxy groups and clusters, 
in order to discuss the ICL properties. 
They simulated a Virgo-like and a Coma-like cluster, and 4 galaxy groups, 
and predict several properties of IC stellar populations. 
\citet{rudick06} have used $N$-body simulations 
(done at both low and high resolutions) of the dynamical evolution 
of galaxy clusters to study the formation and evolution of 
the diffuse ICL component in 3 simulated galaxy clusters. 
They found that the ICL fraction of cluster luminosity increases 
as clusters evolve, reaching $\sim 10 - 15 \%$ in evolved clusters. 

In these numerical studies, there is always a trade-off between having good 
resolution or good statistics. 
\citet{napolitano03,willman04,sommerlarsen05}, and \citet{rudick06} simulate
either one cluster or a few clusters, so even though their simulations have
high resolution, they have poor statistics, in the sense that the cluster(s)
they are simulating might not be representative of the whole cluster population.
At the other extreme,
\citet{murante04} simulate a very large cosmological volume, containing a
statistically significant sample of clusters. Such large simulations however cannot
resolve the scale of dwarf galaxies. Our goal is to have it both ways:
achieving good statistics while resolving the processes responsible
for destroying dwarf galaxies. This is achieved by combining large-scale
cosmological simulations with a semi-analytical treatment of mergers and
tidal disruption.

\subsection{Objectives}
\label{sec-objectives} 

The main objective of the present work is 
to determine if dwarf galaxies in clusters are more prone to destruction
by tides or to destruction by mergers. This determination is then used to predict the contribution 
of dwarf galaxies 
to the origin of intracluster stars in different types of cluster 
environments. 
The DGs in a cluster can be tidally disrupted (by the field of a 
more massive galaxy or by the background halo,) or 
the DGs can be destroyed when they merge with another galaxy. 
The impact of these two destruction mechanisms on the ICL is radically 
different. 
In the case of tidal disruption, the process contributes to IC stars 
in the cluster. 
In the case of merger, the DG is absorbed by a more massive galaxy, 
and there is essentially no contribution to the IC stars. 

For the first part of this project, presented in this paper,
we perform numerical simulations of isolated clusters of galaxies,
in order to examine which method of dwarf galaxy destruction is dominant, 
and how the process depends on environmental factors. 
We identify six possible outcomes for our simulated galaxies: 
(1) the galaxy merges with another galaxy, 
(2) the galaxy is destroyed by the tidal field of a larger galaxy 
but the fragments accrete onto that larger galaxy, 
(3) the galaxy is destroyed by tides 
and the fragments are dispersed in the intracluster medium (ICM), 
contributing to the intracluster light,
(4) the galaxy is destroyed by the tidal field of the background halo, 
(5) the galaxy survives to the present, 
and (6) the galaxy is ejected from the cluster. 

We designed a simple algorithm to follow the evolution of galaxies in 
an isolated cluster. 
The gravitational interaction between galaxies 
is calculated by a direct N-body algorithm. 
The other physical mechanisms 
governing the possible outcomes (mergers, tidal disruption, accretion etc.) 
of the simulated galaxies are treated as ``subgrid physics,'' and
are incorporated in the algorithm using a semi-analytical method. 
In the present work, we use this algorithm to 
simulate the evolution of isolated galaxy clusters, 
i.e. we assume that the cluster has already formed with 
its constituent galaxies in place, and it is neither accreting nor merging. 
Cluster accretion possibly has a finite impact on the evolution of 
dwarf galaxies, 
and on the origin of intracluster stars. 
In a forthcoming paper \citep{bmb08} we will address 
these issues by doing actual cosmological simulations over a statistically
significant volume of the universe. 

The remainder of this paper is organized as follows: 
In \S\ref{sec-numerical} we outline the numerical model for our 
galaxy clusters. 
The methodology of our simulations 
is described in \S\ref{sec-simulations}. 
The results are presented in \S\ref{sec-results} and \S\ref{sec-dependence}. 
We discuss the implications of our main goals in \S\ref{sec-discussion}, 
and give our conclusions in \S\ref{sec-conclusion}. 

\section{THE NUMERICAL METHOD} 
\label{sec-numerical} 

\subsection{The Basic PP Algorithm} 
\label{sec-PPalgo} 

We treat the system as an isolated cluster consisting of $N$ galaxies
of mass $m_i$, radius $s_i$, and internal energy $U_i$,
orbiting inside a background halo of
uncollapsed dark matter and gas. We assume that the halo is spherically
symmetric, and its radial density profile 
$\rho_{\rm halo}^{\phantom2}(r)$ does not
evolve with time (hence, we are neglecting infall motion that would result
from cooling flows). Furthermore, we assume that the halo is stationary:
it does not respond to the forces exerted on it by the galaxies, and 
therefore its center remains fixed at a point that we take to be the origin.

We represent each galaxy by {\it one single particle\/} of mass $m_i$.
The ``radius'' $s_i$ of the galaxy and its ``internal energy'' $U_i$
are internal variables that
only enter in the treatment of the 
subgrid physics described in \S\ref{sec:subgrid} 
below. Our motivation for using 
this approach is the following: To simulate the destruction of dwarf
galaxies by tides, it would seem more appropriate to simulate each
galaxy using many particles. Supposing, however, that it takes at least 100
particles to properly resolve a dwarf galaxy experiencing tidal
destruction, as the galaxies in our simulations cover 3 orders of
magnitude in mass, the most massive ones would be represented
by 100,000 particles. Even though the dwarf galaxies are much more
numerous than the massive ones, the total number of particles would be above
1 million. This raises the following issues:

\begin{itemize}

\item With the use of tree codes, an $N=10^6$-particle simulation
is not considered prohibitive anymore. However (1) our model has
several free parameters, so we have a full parameter-space to study, and
(2) one single cluster is not statistically significant, so for each
combination of parameters we need to perform several simulations.
For this paper, we performed \numberofrunstotal simulations. 
Doing \numberofrunstotal million-particle
simulations would start to be computationally expensive.

\item We could use unequal-mass particles, so that the most massive
galaxies would not be represented by large numbers of particles. This is
usually not a good idea. N-body simulations with particles having
wildly different masses are known to suffer from all sorts of
instability problems, which often require special algorithms to
deal with. The approach we are considering here is more practical.

\item In this paper, we consider isolated clusters. 
In a forthcoming paper \citep{bmb08},
we will present simulations of a cosmological volume containing at least
100 clusters. The number of particles would then reach 100 million, and we
would still need to explore the parameter space. This would be very
computationally expensive. We will solve this problem using single-particle
galaxies combined with a treatment of subgrid physics. The simulations
presented in this paper can be seen as a test-bed for this approach.

\end{itemize}

The relatively small number of particles in our simulations 
(typically less than 1,000) enables us to use a direct, Particle-Particle
(PP) algorithm, which is the simplest of all N-body algorithms. We took
a standard PP code, which evolves a system of $N$ gravitationally 
interacting particles using a second-order Runge-Kutta algorithm. We
modified the original algorithm to include the interaction with the
background halo, and we added several modules to deal with the subgrid
physics. In this modified algorithm, the number of particles $N$ can
vary, as they merge, are destroyed by tides, or escape the cluster. 


\subsection{Gravitational Interactions}
\label{sec:gravity}

The acceleration of particle $i$ (or galaxy $i$) is given by
\begin{equation}
\label{gravity}
{\bf a}_i=-G\sum_{j\ne i}{m_j({\bf r}_i-{\bf r}_j)\over
\bigl(|{\bf r}_i-{\bf r}_j|^2+\epsilon^2\bigr)^{3/2}}
-{GM_{\rm halo}(r_i){\bf r}_i\over(r_i^2+\epsilon^2)^{3/2}}\,,
\end{equation}

\noindent where ${\bf r}_i$ and ${\bf r}_j$ are the positions of
particles $i$ and $j$, respectively, $m_j$ is the mass of particle $j$,
$M_{\rm halo}(r_i)$ is the mass of the background halo inside $r=r_i$,
$G$ is the gravitational constant, and $\epsilon$ is the softening length.
This assumes that the background cluster halo is spherically symmetric and
centered at the origin. In our PP algorithm, this expression is evaluated
directly, by summing over all particles $j\neq i$. The softening length
$\epsilon$ is chosen to be smaller than the initial radius of the smallest 
galaxies (see \S3.2 below for the determination of the initial radius).

We evolve the system forward in time using a second-order Runge Kutta
algorithm. The timestep $\Delta t$ is calculated using
\begin{equation}
\Delta t=\min_i\,(\Delta t)_i\,,\quad
(\Delta t)_i=\min\left[{\epsilon\over|{\bf v}_i|},
\left({\epsilon\over|{\bf a}_i|}\right)^{1/2}\right]\,,
\end{equation}

\noindent where ${\bf v}_i$ is the velocity
of particle $i$, and we take 
the smallest value of $(\Delta t)_i$ to be the timestep $\Delta t$.

\subsection{The Cluster Halo Density Profile}
\label{sec-haloRho}

We consider two different types of density profile of the background 
halo of a cluster, $\rho_{\rm halo}(r)$: the $\beta$ profile, 
and the NFW profile.

In the first case, we assume that the dark matter in the 
background halo follows a 
similar density distribution as the observed intracluster gas. 
A single $\beta$-model (isothermal) density profile is used for the gas 
(e.g. \citealt{king62, cavaliereetal76, makinoetal98}), 
\begin{equation}
\label{rho-beta}
\rho_{\rm gas}\left(r\right)
=\rho_0^{\phantom2}\left[1+\left(r/r_c\right)^2 \right]^{-3\beta/2}, 
\end{equation} 
where, $\rho_0$ is the central density, and $r_c$ is the core radius. 
The values of $\rho_0^{\phantom2}$, $r_c$ and $\beta$ are taken from 
\citet{piffarettietal06}, 
which gives the gas density parameters for 16 nearby clusters. 
The halo density is then obtained by scaling the gas density with 
the universal ratio of matter (dark + baryonic) to baryons, 
$\rho_{\rm halo}^{\phantom2} 
=\rho_{\rm DM}^{\phantom2}+\rho_{\rm gas}^{\phantom2}= 
\rho_{\rm gas}^{\phantom2}\Omega_M/\Omega_b$, where $\Omega_M$ and
$\Omega_b$ are the present matter (baryons + dark matter) density parameter 
and baryon density parameter, respectively. 
This assumes that the cluster baryon mass fraction follows the 
cosmic value of $\Omega_b/\Omega_M$, which is expected to be generally true 
\citep[e.g.,][]{white93, allen02, ettori03}, 
although precise estimations of cluster baryon content have shown deviations 
from the universal value \citep[][and references therein]{gonzalez07}. 

In a second case, we consider that the distribution of gas and 
dark matter in the background halo both follow
analytical models of the dark matter density 
having a functional form 
\begin{equation}
\label{rho-anal}
\rho_{\rm DM}^{\phantom2} \left(r\right)=\frac{\rho_s} {\left(r/r_s\right) 
\left(1+r/r_s\right)^2}
\end{equation} 
\citep{NFW97}.
Here, $\rho_s$ is a scaling density, and $r_s$ is a scale length. 
The NFW profile is often parametrized in terms of a concentration
parameter $c$. The parameters $\rho_s$ and $r_s$ are then given by
\begin{eqnarray}
\rho_s&=&{200c^3\rho_{\rm crit}(z)/3\over
{\ln\left(1\!+\!c\right)-c/\left(1\!+\!c\right)}}
={25H^2(z)c^3/\pi G\over
{\ln\left(1\!+\!c\right)-c/\left(1\!+\!c\right)}},
\\
r_s&=&{r_{200}^{\phantom2}\over c}\,,
\end{eqnarray}

\noindent where
$\rho_{\rm crit}(z)=3H^2(z)/8\pi G$
 is the critical density at formation redshift $z$, and
$r_{200}$, the virial radius, is the radius of a sphere whose 
mean density is $200 \rho_{\rm crit}$ (200 times the critical density 
of the Universe at the epoch of formation). After scaling, the halo
density profile is
$\rho_{\rm halo}^{\phantom2} 
= \rho_{\rm DM}^{\phantom2}\Omega_M/(\Omega_M-\Omega_b)$.

Once we have chosen a particular density profile,
the density is integrated to get the background cluster halo mass as 
\begin{equation}
\label{mhalo}
M_{\rm halo}(r)=\int_0^r 4 \pi x^2 \rho_{\rm halo}^{\phantom2}(x) dx. 
\end{equation}

\noindent This is the mass that enters in the last term of 
equation~(\ref{gravity}). Since the density profiles we consider do not
have an outer edge where $\rho_{\rm halo}^{\phantom2}=0$, 
we truncate the cluster background halo at
a maximum halo radius $R^{\max}_{\rm halo}=5\,{\rm Mpc}$. Equation~(\ref{mhalo})
is then solved numerically, to build an interpolation table for
$r$ in the range $[0,R^{\max}_{\rm halo}]$ that is then used by the code.

\subsection{The Subgrid Physics}
\label{sec:subgrid} 

As mentioned in \S\ref{sec-objectives}, 
there can be six possible physical outcomes for our simulated cluster galaxies. 
In the following subsections, 
we describe the associated subgrid physics for each mechanism 
we use in our simulations. 
The possible outcomes are: 
(1) the galaxy merges with another galaxy (\S\ref{sec-merger}), 
(2) the galaxy is destroyed by the tidal field of a larger galaxy 
but the fragments accrete onto that larger galaxy (\S\ref{sec-tideAccr}), 
(3) the galaxy is destroyed by tides of a larger galaxy 
and the fragments are dispersed in the intracluster medium (\S\ref{sec-tideICL}), 
(4) the galaxy is destroyed by the tidal field of the background halo 
(\S\ref{sec-tideICL}), 
(5) the galaxy survives to the present (i.e., it is not destroyed by any process), 
and (6) the galaxy is ejected from the cluster (\S\ref{sec-eject}). 
We describe our approach of simulating galaxy harassment  in \S\ref{sec-harass}. 

\subsubsection{Encounter: Merger} 
\label{sec-merger}

We simulate a pair of galaxies colliding 
(or synonymously, having an encounter) 
and the further consequences (e.g., merging) in the following way. 
An encounter is accounted for when two galaxies $i$ and $j$,
of radii $s_i$ and $s_j$, overlap such that 
the center of the galaxy $j$ is inside the galaxy $i$. 
Numerically the criterion is $r_{ij} < s_i$, 
where, $r_{ij} = |{\bf r}_i-{\bf r}_j|$ is the distance 
between the centers of the galaxies. 
If ${\bf v}_i$ and ${\bf v}_j$ are the velocities of galaxies $i$ and $j$, 
the center of mass velocity of the pair is 
${\bf v}_{\rm cm}=\left(m_i{\bf v}_i+m_j{\bf v}_j\right)/\left(m_i+m_j\right)$.
The kinetic energy in the center-of-mass rest frame is
\begin{equation}
K_{ij}=\frac{1}{2} m_i|{\bf v}_i-{\bf v}_{\rm cm}|^2+\frac{1}{2}m_j
|{\bf v}_j-{\bf v}_{\rm cm}|^2. 
\end{equation} 
The gravitational potential energy of the pair is
\begin{equation}
W_{ij} = - \frac{G m_i m_j}{r_{ij}}.
\end{equation} 

Even though we are treating each galaxy as a single particle, in reality
a galaxy is a gravitationally bound system with an internal kinetic 
energy and a potential
energy, and these energies must be included in the total energy of the
interacting pair of galaxies.
Considering a galaxy as a bound virialized system its internal energy is 
\begin{equation}
\label{internal}
U_i=U_{\rm potential}+U_{\rm kinetic}={U_{\rm potential}\over2}
=-\frac{\zeta G m_i^2}{2 s_i}, 
\end{equation} 
where $\zeta$ is a geometrical factor which depends on 
the mass distribution in the galaxies. Throughout this paper,
we assume $\zeta=1$ (see Appendix A).

We then compute the total energy of the galaxy pair 
(in the center of mass frame) as 
\begin{equation}
\label{E-pair} 
E_{ij} = K_{ij}  + W_{ij} + U_i + U_j.
\end{equation} 
If $E_{ij} \leq 0$, i.e., the system is bound, 
we then allow the galaxies to merge to form a single
galaxy of mass $m_{\rm merge}=m_i + m_j$.
To compute its radius, we assume that energy is conserved, 
hence the total energy $E_{ij}$ in the center-of-mass rest frame is all
converted into the internal energy of the merger remnant.
Its radius is then computed from equation~(\ref{internal}),
\begin{equation}
\label{s_merged}
s_{\rm merged}=-{\zeta Gm_{\rm merge}^2\over 2U_{\rm merge}}
={\zeta G (m_i+m_j)^2\over2|E_{ij}|}\,.
\end{equation}
The position and velocity of this merged object are set to 
those of the center-of-mass values of the galaxy pair before merger, 
in order to conserve momentum. 

\subsubsection{Encounter: Galaxy Harassment} 
\label{sec-harass}

In a high-speed encounter the two interacting galaxies 
come into contact for a brief amount of time. 
The galaxies might survive a merger or tidal disruption, 
but the encounter adds some internal energy into them,
making them less bound. We refer to this process as {\it Galaxy Harassment}.
This process has been originally suggested as a possible explanation for
the origin of the morphology-density relation in clusters \citep{moore96}.
We incorporate galaxy harassment in our algorithm by increasing the radius 
(or the internal energy) 
of a galaxy 
when it experiences a non-merger encounter. 
This enlargement makes a galaxy more prone to tidal
disruption at a next encounter.

In equation~(\ref{E-pair}), if $E_{ij} > 0$, i.e., the system is not bound, 
then the galaxies in our simulation do not merge in the collision. 
Rather a part of the kinetic energy of the galaxies is converted into 
internal energy, making the collision inelastic. We assume that an
equal amount of energy is transfered to each galaxy.
Denoting the energy transfered as $\Delta E$, 
the kinetic energy of the pair decreases by $\Delta E$, while the internal
energy of each galaxy increases by $\Delta E/2$. 
We assume
\begin{equation}
\label{harass} 
{\Delta E\over2}=\eta\min\left(|U_i|,|U_j|\right)\,, 
\end{equation} 
where $\eta$ is an energy transfer efficiency whose 
value is taken as $\eta=0.2$. The internal energies of the two
galaxies after the encounter are 
$U_i^{\rm after}=U_i^{\rm before}+\Delta E/2$ and
$U_j^{\rm after}=U_j^{\rm before}+\Delta E/2$, respectively.
By choosing $\eta<1$, we are ensuring that the internal energy of each
galaxy
remains negative, that is, the transfer of energy does not unbind the
galaxies.
We recalculate the velocities ${\bf v}_i$ and ${\bf v}_j$ after 
collision while conserving momentum, assuming that 
only the magnitudes of velocity change, the directions remaining the same. 
We recalculate the size of each galaxy in the pair using 
equation~(\ref{s_merged}),
\begin{equation} 
\label{s_ij_after} 
s_i^{\rm after}= \frac{\zeta G m_i^2}{2 |U_i^{\rm after}|}\,,\qquad
s_j^{\rm after}= \frac{\zeta G m_j^2}{2 |U_j^{\rm after}|}. 
\end{equation} 

While allowing a size increase of the galaxies 
according to equations~(\ref{s_merged}) and (\ref{s_ij_after}), 
we also considered a size cutoff. 
We assumed that the galaxies could grow only up to a maximum size 
given by the size of the largest galaxy at the beginning of the simulation. 

\subsubsection{Tidal Disruption: Intracluster Stars} 
\label{sec-tideICL} 

\begin{figure}
\begin{center}
\includegraphics[width=3.5in]{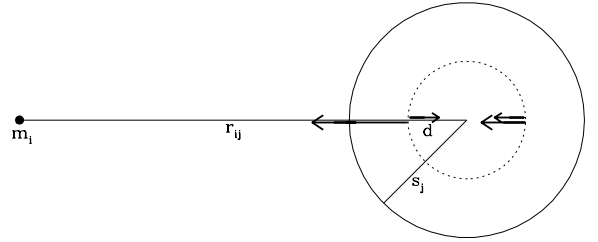}
\caption{Calculation of the effects of tides caused by 
a galaxy of mass $m_i$ on a galaxy of mass $m_j$ and radius $s_j$. 
The two largest arrows show the gravitational accelerations caused by galaxy $i$; 
the two smallest arrows show the accelerations caused by galaxy $j$.
See \S\ref{sec-tideICL} for details.}
\label{tides_schem}
\end{center}
\end{figure}

We consider two possible sources of external gravitation for the tidal 
disruption of a galaxy $j$: 
another galaxy $i$, or the background cluster halo. 
The tidal force on a galaxy due to the gravitational field of the external 
source is meaningful only if the galaxy lies entirely on one side of the 
external source, 
when the tides are directed radially outwards tending to tear apart 
the galaxy. Our calculation of the tidal field caused by
a galaxy $i$ of mass $m_i$ is illustrated in 
Figure~\ref{tides_schem}. The galaxies~$i$ and $j$ are separated by a 
distance $r_{ij}$.
We calculate the resultant fields between two diametrically opposite points 
inside galaxy $j$, located at a radial distance $d\leq s_j$ 
along the line joining the centers of the two galaxies. 
The two small and two large arrows in Figure~\ref{tides_schem} 
indicate the gravitational field (or acceleration) at the opposite points 
caused by galaxy $j$ (self-gravity) and 
by galaxy~$i$ (external source of gravitation), respectively.
The magnitude of the tidal field is
given by the difference between the gravitational field caused by galaxy $i$
at the two opposite points,\footnote{This reduces to the well-known form
$a_{\rm tide}^{\rm galaxy}\propto d/r_{ij}^3$ in the limit $d\ll r_{ij}$,
but we do not make this approximation here.}
\begin{equation} 
\label{F-tide} 
a_{\rm tide}^{\rm galaxy} = \frac{G m_i}{\left(r_{ij} - d\right)^2} 
                - \frac{G m_i}{\left(r_{ij} + d\right)^2}. 
\end{equation} 

\noindent 
The gravitational field caused by galaxy $j$ 
(two small arrows in Figure~\ref{tides_schem}) is directed 
radially inwards and acts opposite to the tides,
tending to keep the galactic mass inside radius $d$ intact. 
The difference between that self-gravitational field 
at the two opposite points is 
\begin{equation} 
\label{F-grav} 
a_{\rm grav}^{\phantom2} = \frac{2 G m_j(d)}{d^2}, 
\end{equation} 

\noindent where $m_j(d)$ is the mass of galaxy $j$ inside radius $d$.
When $a_{\rm tide}^{\rm galaxy}=a_{\rm grav}^{\phantom2}$, then the tides will
exceed self-gravity at radii larger than $d$, while self-gravity
will exceed the tides at smaller radii.
Thus the layers of galactic mass 
located between radii $d$ and $s_j$ would become
unbound, while the ones located inside radius $d$ would remain bound.
Hence, the galaxy would be partly disrupted. In our code, we simplify
things by using an ``all-or-nothing'' approach. A galaxy is either
totally disrupted,
or not disrupted at all. We consider that a galaxy 
is disrupted if half of its mass or more becomes unbound. If we assume
an isothermal sphere density profile (as in Appendix A), then
the half-mass radius is given by $d=s_j/2$. This is the value of
$d$ we use in equations~(\ref{F-tide}) and (\ref{F-grav}). The criterion
of tidal destruction then becomes 
$a_{\rm tide}^{\rm galaxy}(d)\geq a_{\rm grav}^{\phantom2}(d)$,
with $d=s_j/2$.

We also consider tidal disruption by the background cluster halo, but only if
$r_j>s_j$ (that is, the galaxy does not overlap with the center of the halo).
The magnitude of tidal field due to the cluster halo is 
\begin{equation} 
\label{tides-halo}
a_{\rm tide}^{\rm halo} = 
\frac{GM_{\rm halo}(r_j - d)} {(r_j - d)^2} 
- \frac{GM_{\rm halo}(r_j + d)}{(r_j + d)^2}\,,
\end{equation} 
 
\noindent where $M_{\rm halo}(r)$ is given by equation~(\ref{mhalo}).
If $a_{\rm tide}^{\rm halo}(d)\geq a_{\rm grav}(d)$, with $d=s_j/2$,
galaxy $j$ is tidally destroyed by the gravitational field of the halo. 

When galaxy $j$ is considered to have been tidally destroyed 
by another galaxy $i$, the fragments of the disrupted galaxy 
might accrete onto galaxy $i$ (\S\ref{sec-tideAccr}), 
or they might be dispersed into the ICM when $E_{ij} > 0$ 
[$E_{ij}$ being the total energy of the galaxy pair given by eq.~(\ref{E-pair})]. 
For tidal destruction by the cluster halo the disrupted fragments 
are always dispersed into the ICM. 
In both cases, the destroyed galaxy is removed from the list of existing particles. 
The code keeps track of the amount of mass added to the ICM 
(in the form of IC stars) by tidal disruption. This quantity is initialized
to zero at the beginning of the simulation, and every time a galaxy is
tidally destroyed with its fragment dispersed, the mass of that
galaxy is added up to the mass of IC stars. 

\subsubsection{Tidal Disruption: Accretion} 
\label{sec-tideAccr}

We consider a possibility of accretion of the fragments of a 
tidally disrupted galaxy onto the galaxy causing the tides. 
This happens for the case of tidal disruption due to galaxies only
(if the disruption is caused by the background cluster halo, 
the fragments are always dispersed as IC stars). 
This situation occurs when the conditions 
$a_{\rm tide}^{\rm galaxy}>a_{\rm grav}^{\phantom2}$ and $E_{ij} \leq 0$ are 
both satisfied (see \S\ref{sec-tideICL}).
The tidally disrupted galaxy accretes onto the more massive galaxy. 
The mass of the bigger galaxy increases from $m_i$ to $m_i+m_j$. 
Thus a tidal disruption 
followed by accretion is similar to a merger (\S\ref{sec-merger}), 
but these events are counted separately.

\subsubsection{Ejection}
\label{sec-eject}

When a galaxy ventures at distances larger than the cluster halo 
truncation radius $R^{\max}_{\rm halo}$ (see \S\ref{sec-haloRho}), 
we consider that this galaxy has escaped from the cluster, 
and we remove it from the list. If we kept that galaxy, it might 
eventually return to the cluster. But in reality the universe contains 
many clusters, and a galaxy that moves sufficiently far away from one cluster
will eventually feel the gravitational influence of other clusters, 
something that our algorithm, which simulates an isolated cluster,
does not take into account. As we shall see, the
ejection of galaxies from a cluster is quite uncommon in our simulations. 

\section{THE SIMULATIONS} 
\label{sec-simulations} 

\subsection{Cosmological Model}

We consider a $\Lambda$CDM model with the 
present matter density parameter, $\Omega_M=0.241$,
baryon density parameter, $\Omega_b=0.0416$, cosmological
constant, $\Omega_\Lambda=0.759$, Hubble constant,
$H_0=73.2\rm\,km\,s^{-1}Mpc^{-1}$ ($h=0.732$),
primordial tilt,
$n_s=0.958$, and CMB temperature, $T_{\rm CMB}=2.725$ K, consistent with the
results of {\sl WMAP3} \citep{spergeletal07}. Even though the simulations 
presented in this paper are not ``cosmological'' (we simulate isolated,
virialized clusters), the particular choice of cosmological model enters the picture twice:
in the determination of the radii of galaxies (see next section), and in the
calculation of the elapsed time between the initial and final redshifts of the
simulation. 

In each simulation a cluster is evolved from $z=1$ 
to the present ($z=0$). We assume that the cluster will not experience any
major merger during this period, and that, therefore, it is a good
approximation to treat it as isolated. For our $\Lambda$CDM model,
this represents a total evolutionary time of 7.63~Gyr.

\subsection{Initial Conditions}
\label{sec-initial}

\begin{figure*}
\begin{center}
\includegraphics[width=6in]{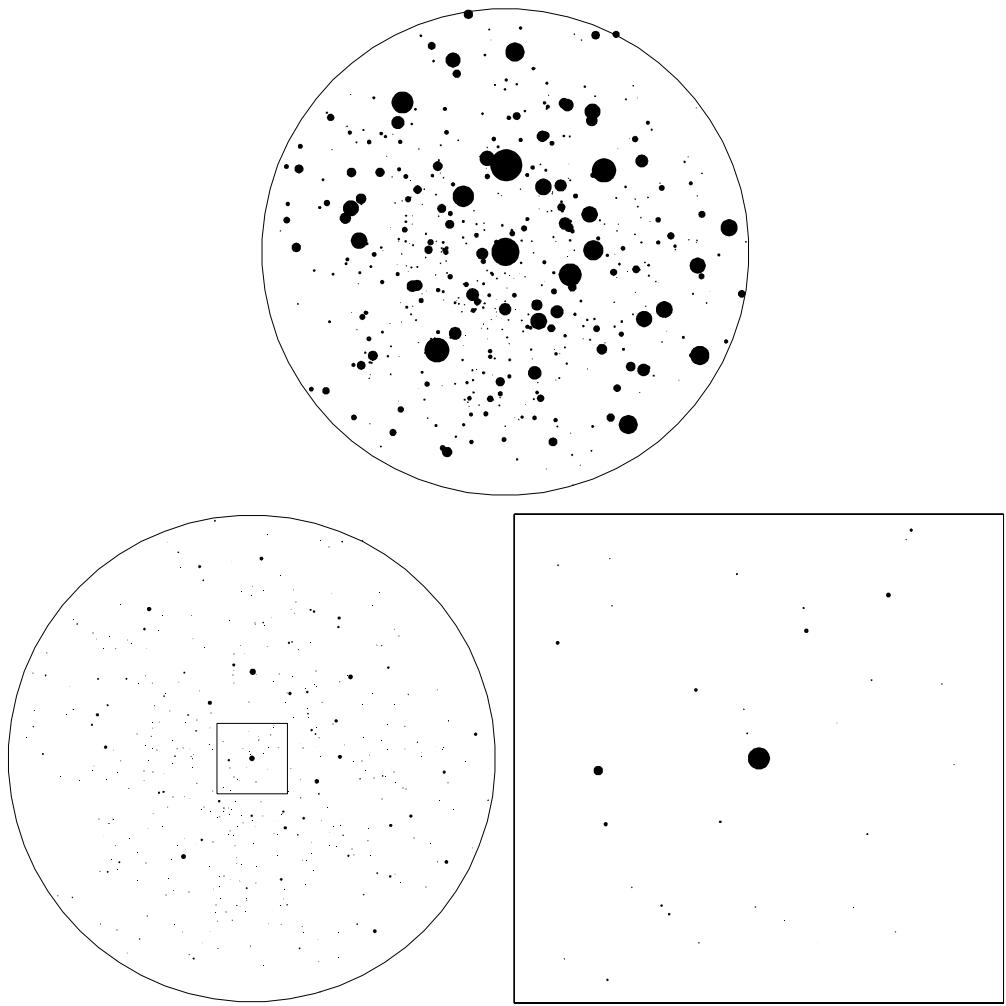}
\caption{Initial conditions for run A12. Top panel:
initial conditions at $z=1$. The solid circles indicate the
virial radii of galaxies. The large circle is the maximum distance 
$r=3R_0=2.08\,\rm Mpc$
from the cluster center. Lower left panel: same as top panel, 
with symbols rescaled to optical diameter of real galaxies. 
Bottom right panel: enlargement of the central $(0.6\,{\rm Mpc})^2$, 
(box on lower-left panel).}
\label{init_A12}
\end{center}
\end{figure*}

To set the initial conditions of our simulations, 
we need to determine the initial mass $m$, radius $s$, 
position ${\bf r}$, and velocity ${\bf v}$ of each galaxy. 
To determine the mass, we first assume that the luminosities
of galaxies are distributed according to the 
Schechter luminosity function \citep{schechter76},

\begin{equation}
\label{schechter}
\phi(L)dL=\phi^*\biggl({L\over L^*}\biggr)^\alpha e^{-L/L^*}{dL\over L^*}\,.
\end{equation}

\noindent Here we use $L^*= 3.097\ee{10} L_\odot$ 
(corresponding to absolute magnitude $M_{b_J}^*=-20.07$), 
and $\alpha=-1.28$, which is appropriate
for galaxies located in clusters \citep{deproprisetal2003}. 
This luminosity function spans over $-22.5 < M_{b_J} < -15$. 
While it might be reasonable
to assume fixed values of $L^*$ and $\alpha$, the value of $\phi^*$
most likely varies amongst clusters. So we normalize 
equation~(\ref{schechter}) by imposing that, in each cluster, 
there are $N_0$ galaxies with luminosities $L>L_0$. 
We use $N_0 = 25$, and $L_0 = 0.2 L^*$ (corresponding to $M_b=-19$) 
\citep{lewis02}.
We select the luminosities using a Monte Carlo rejection method. 
We then assume a constant mass-to-light ratio 
$\Upsilon=193~h~M_\odot/L_\odot = 73$ \citep{brainerdetal03}, 
and convert the luminosities to masses. 
The Schechter function spans up to a maximum mass 
$M_{\rm max}=220 \ee{11} \msun$. 
To generate the dwarf galaxies, the same Schechter function 
is extrapolated up to a minimum mass $M_{\rm min}=1 \ee{9} \msun$. 

We take the radius $s$ of each galaxy to be equal to 
the virial radius $r_{200}$ (radius containing matter with 200 times 
the mean density of the Universe at the epoch of galaxy formation) 
corresponding to the galaxy mass $m=M_{200}$ using 
\begin{equation}
\label{m200} 
M_{200} = \frac{800\pi}{3} r_{200}^3\bar\rho\left(1+z_{\rm coll}\right)^3. 
\end{equation} 
Here, $\bar\rho=\rho_{\rm crit}\Omega_M=3H_0^2\Omega_M/8\pi G$ 
is the mean matter density in the present universe, and 
$z_{\rm coll}$ is the redshift of collapse when the galaxy formed.
To obtain $z_{\rm coll}$, we use a simple spherical collapse model.
First, by filtering the power spectrum for our $\Lambda$CDM model,
we calculate the standard deviation $\sigma(m)$ of the linear density
contrast $\delta=(\rho-\bar\rho)/\bar\rho$ at the mass scale $m$.
The distribution of the values of $\delta$ is then given by a Gaussian,
\begin{equation} 
{\cal P} (\delta) \propto \exp \left(- \frac{\delta^2}{2 \sigma^2} \right). 
\end{equation} 
We pick randomly a present density contrast 
$\delta_0=\delta(z=0)$ from this distribution,
using a Monte Carlo rejection method, 
and solve the following equation to get the collapse redshift 
$z_{\rm coll}$,

\begin{equation}
\Delta_c=\delta_0{\delta_+(z_{\rm coll})\over\delta_+(0)}\,,
\end{equation}

\noindent where $\delta_+(z)$ is the linear growing mode
(for $\Lambda\neq0$ models, see, e.g., \citealt{martel91}).
Here, $\Delta_c=3(12\pi)^{2/3}/20=1.686$ 
is the overdensity predicted by linear theory at recollapse. We solve
this equation numerically for $z_{\rm coll}$, and substitute the solution in
equation~(\ref{m200}), which we then solve to get the radius $s=r_{200}$.

To determine the locations of galaxies inside a cluster, 
we assume that their distribution is isotropic (in a statistical sense). 
We can therefore choose the spherical coordinates
$(\theta,\phi)$ of each galaxy randomly, using $\phi=2\pi X_\phi$,
and $\cos\theta=2X_\theta-1$, where $X_\phi$ and $X_\theta$ are
random numbers drawn from a uniform deviate between 0 and 1. We still
need to determine the radial coordinate $r$. Using the CNOC cluster
survey, \citet{carlbergetal97} showed that the radial mass density $\rho(r)$ of 
matter and the radial number density $\nu(r)$ of galaxies are roughly proportional 
to each other, where both $\rho(r)$ and $\nu(r)$ are approximated by NFW profiles. 
\citet{girardi98} found that the halo mass follows 
the galaxy distribution in clusters, 
using a $\beta$-model for the halo/galaxy volume density profile. 
We assume that this proportionality holds for all clusters, and we 
generalize it to to all the density profiles we use. 
Thus, the assumed background halo mass density profile 
[eqs.~(\ref{rho-beta}) and (\ref{rho-anal})] gives us $\nu(r)$. We can then select
the initial distances $r$ from the cluster center using again
a Monte Carlo rejection method.
Since the masses and locations of the galaxies have been determined
separately, we need to pair them, that is,
for each selected location, to decide which galaxy goes there.
We do expect the most massive galaxies to reside near the center
of the cluster. However, low-mass galaxies are not all located
at large radii, and some of them might be located in the central
region of the cluster as well. Indeed, if the galaxies in the
central region were all massive, it would be impossible to reproduce
the desired number density profile $\nu(r)$ {\it and\/} not have the
galaxies overlap.

To prevent any overlap, we locate the galaxies as follows: we first
position the most massive galaxy at the center of the cluster. Then we
locate the next 7 most massive galaxies between radii $R_0$ and $2R_0$,
where $R_0$ is 3 times the radius of the most massive galaxy. We then
locate the next 19 most massive galaxies between radii $2R_0$ and $3R_0$. Finally,
the remaining galaxies are located randomly between radii $0$ and $3R_0$.
During the process we check that the galaxies do not overlap, by ensuring
that the distance between the edges of two galaxies 
is greater than the radius of the larger galaxy, 
i.e., $r_{ij} - s_i - s_j > \max (s_i, s_j)$. 
In the process of locating a galaxy, if this criterion is not satisfied, 
we simply reject that location and generate a new one. 

After assigning the masses, radii, and positions of all the galaxies, 
we determine the velocity of each galaxy. We consider the velocity 
a galaxy would have if it were in a perfect circular orbit at radius $r$, 
\begin{equation}
v_{\rm circ}^{\phantom1}(r)=\left[{G\over r}
\left(M_{\rm halo}(r)+\sum_{j,\>r_j<r}m_j\right)\right]^{1/2}\,, 
\end{equation} 

\noindent where the sum only includes galaxies inside  radius $r$. 
The norm of the velocity is chosen by giving a random 10\% 
deviation to the circular velocity, i.e., 
$v=v_{\rm circ}^{\phantom1}(1+ 0.1 X_v)$, 
where $X_v$ is a random number between $-1$ and $1$. 
For the direction of the velocity, we follow a similar random angle 
generation technique as we did for the positions of galaxies. 

\begin{deluxetable*}{cccccccccc}
\tabletypesize{\scriptsize}
\tablecaption{Series of Simulations}
\tablewidth{0pt}
\tablehead{
\colhead{Series} & \colhead{Runs} &
\colhead{$\alpha_{\rm start}^{\phantom1}$} & 
\colhead{profile} &
\colhead{$\beta$} &
\colhead{$\rho_0^{\phantom1},\rho_s^{\phantom1}\,[\rm g\,cm^{-3}]$} &
\colhead{c} & 
\colhead{$r_c^{\phantom1},r_s^{\phantom1}\,[{\rm kpc}]$} & 
\colhead{cD} &
\colhead{Harassment}
}
\startdata
 A & 16 & $-1.28$ & $\beta$-Virgo   & 0.33 & $8.14\times10^{-26}$ &$\cdots$ & 
3 & $\times$ & $\times$ \cr
 B & 17 & $-1.28$ & $\beta$-Virgo   & 0.33 & $8.14\times10^{-26}$ &$\cdots$ &  
3 & $\times$ & $\surd$  \cr
 C & 17 & $-1.36$ & $\beta$-Virgo   & 0.33 & $8.14\times10^{-26}$ &$\cdots$ &  
3 & $\times$ & $\surd$  \cr
 D & 16 & $-1.36$ & $\beta$-Virgo   & 0.33 & $8.14\times10^{-26}$ &$\cdots$ &  
3 & $\surd$  & $\surd$  \cr
\noalign{\hrule}
 E & 16 & $-1.36$ & $\beta$-Perseus & 0.53 & $7.27\times10^{-26}$ &$\cdots$ & 
28 & $\times$ & $\surd$  \cr
 F & 16 & $-1.36$ & $\beta$-Perseus & 0.53 & $7.27\times10^{-26}$ &$\cdots$ & 
28 & $\surd$  & $\surd$  \cr
\noalign{\hrule}
 G &  10 & $-1.28$ & NFW & $\cdots$ & $2.35\times10^{-25}$& 5 & 
200 & $\times$ & $\surd$  \cr
 H &  14 & $-1.31$ & NFW & $\cdots$ & $2.35\times10^{-25}$& 5 & 
200 & $\times$ & $\surd$  \cr
 I &  10 & $-1.31$ & NFW & $\cdots$ & $2.35\times10^{-25}$& 5 & 
200 & $\surd$  & $\surd$  \cr
\enddata
\label{series}
\end{deluxetable*}

Figure~\ref{init_A12} illustrates the initial conditions for one of our simulations
(run A12). The top panel shows the cluster at $z=1$. The large circle represents
the maximum distance $3R_0$ within which the galaxies are located initially. 
Each dot represents a galaxy, with the most massive one located in the center. 
Even though massive galaxies tend to be larger, 
there is no direct correspondence between the masses and 
radii because of the dependence on $z_{\rm coll}$ in equation~(\ref{m200}), 
whose determination involves a Monte Carlo method. 

Visually, this looks quite different from the optical pictures of actual clusters
like Virgo. This is because each dot has a radius $s$ equal to the virial 
radius $r_{200}$, that can exceed the optical radius by an order of magnitude.
In the bottom left panel of Figure~\ref{init_A12}, we show the same cluster, 
with all the dots rescaled in size so that the angular diameter of the central galaxy 
is equal to $8.3'$ at a distance of $16.8\,\rm Mpc$, 
which is the observed optical diameter of M87. The bottom right panel
shows a zoom-in of the central cluster region. 
It looks qualitatively similar to pictures of the central region of Virgo.

\section{RESULTS} 
\label{sec-results}

We started by performing 
9 series of simulations, for a total of \numberofruns simulations. 
Table 1 summarizes the characteristics 
of each series. The first 2 columns show the series name and the number of 
runs, respectively. The slope of the Schechter luminosity function
at $z=1$ (used to generate the initial conditions) is listed in column 3.
Columns 4 to 8 give the characteristics and relevant parameter values 
of the background cluster halo profile. 
Columns 9 and 10 indicate respectively whether a cD
galaxy was included in the cluster simulation, 
and whether galaxy harassment was included as part of the subgrid physics. 

\subsection{Series A: Initial Simulations}

\begin{deluxetable*}{cccccccccccc}
\tabletypesize{\scriptsize}
\tablecaption{Simulations for Series A}
\tablewidth{0pt}
\tablehead{
\colhead{Run} & \colhead{$M_{\rm total}[10^{11}M_\odot]$} 
& \colhead{$N_{\rm total}$} &
\colhead{$N_{\rm merge}$} & \colhead{$N_{\rm tides}^{\rm gal}$} & 
\colhead{$N_{\rm accr}$} & \colhead{$N_{\rm tides}^{\rm halo}$} & 
\colhead{$N_{\rm eject}$} &
\colhead{$f_{\rm surv}^{\phantom1}$} &
\colhead{$f_{\rm ICS}^{\phantom1}$} &
\colhead{$\alpha_{\rm start}^{\phantom1}$} & 
\colhead{$\alpha_{\rm end}^{\phantom1}$} 
}
\startdata
 A1 &  855.0 & 440 & 137 &  54 &  6 & 0 & 1 & 0.550 & 0.225 & $-1.26$ & $-1.23$ \cr
 A2 &  862.0 & 702 & 343 & 182 & 24 & 0 & 1 & 0.217 & 0.490 & $-1.28$ & $-1.12$ \cr
 A3 & 1100.4 & 480 & 126 &  43 &  7 & 0 & 0 & 0.633 & 0.136 & $-1.28$ & $-1.24$ \cr
 A4 &  831.0 & 530 & 165 &  67 & 11 & 0 & 0 & 0.542 & 0.148 & $-1.26$ & $-1.15$ \cr
 A5 &  980.4 & 459 & 175 &  58 &  8 & 0 & 1 & 0.473 & 0.128 & $-1.25$ & $-1.18$ \cr
 A6 &  967.8 & 640 & 230 &  74 &  5 & 0 & 0 & 0.517 & 0.157 & $-1.29$ & $-1.23$ \cr
 A7 &  720.0 & 579 & 263 & 114 &  7 & 0 & 0 & 0.337 & 0.359 & $-1.28$ & $-1.20$ \cr
 A8 &  757.8 & 435 & 200 &  66 & 51 & 0 & 1 & 0.269 & 0.320 & $-1.27$ & $-1.26$ \cr
 A9 &  925.6 & 514 & 223 &  83 & 12 & 0 & 1 & 0.379 & 0.245 & $-1.28$ & $-1.17$ \cr
A10 &  858.9 & 525 & 204 &  54 &  8 & 0 & 0 & 0.493 & 0.257 & $-1.31$ & $-1.24$ \cr
A11 &  880.2 & 547 & 224 &  68 &  9 & 0 & 1 & 0.448 & 0.257 & $-1.33$ & $-1.25$ \cr
A12 &  826.1 & 548 & 255 &  72 & 15 & 0 & 1 & 0.374 & 0.225 & $-1.29$ & $-1.18$ \cr
A13 & 1041.5 & 431 & 166 &  51 &  6 & 0 & 0 & 0.483 & 0.275 & $-1.27$ & $-1.25$ \cr
A14 &  957.3 & 486 & 174 &  52 &  4 & 0 & 0 & 0.527 & 0.261 & $-1.29$ & $-1.22$ \cr
A15 &  860.5 & 520 & 160 &  64 & 15 & 0 & 1 & 0.538 & 0.260 & $-1.26$ & $-1.18$ \cr
A16 &  858.0 & 483 & 255 &  72 &  8 & 0 & 0 & 0.306 & 0.323 & $-1.28$ & $-1.19$ \cr
\enddata
\label{seriesa}
\end{deluxetable*}

We performed an initial series of 16 simulations, using for the background halo
a $\beta$-profile with $\beta=0.33$, a core radius $r_c=3\,\rm kpc$, and
a central density $\rho_0=8.14\times10^{-26}\,\rm g\,cm^{-3}$, 
which is appropriate for a cluster like Virgo \citep{piffarettietal06}. 
For this series, we did not include galaxy harassment. 
Our results are shown in Table~\ref{seriesa}. 
It shows the run number (column 1), and, at the beginning of the run, 
the total mass $M_{\rm total}$ in galaxies, in units
of $10^{11}M_\odot$ (column 2), 
the number of galaxies $N_{\rm total}$ (column 3), 
and the Schechter luminosity function [eq.~(\ref{schechter})]
exponent $\alpha_{\rm start}^{\phantom1}$ (column 11). 
This exponent was obtained by performing a numerical
fit to the distribution of galaxy masses. Because the masses were
determined from a Monte Carlo rejection method, 
the exponent can differ from the intended value $\alpha=-1.28$ 
in equation~(\ref{schechter}) and listed in Table~\ref{series}, 
but the deviations are small. Averaging over all runs, we get
\begin{equation}
\alpha_{\rm start}^{\phantom1}=-1.280\pm0.020\,.
\end{equation}

Columns $4-8$ in Table~\ref{seriesa} show 
the number of galaxies $N_{\rm merge}$ destroyed by mergers, 
the number of galaxies $N_{\rm tides}^{\rm gal}$ destroyed by tides caused
by a massive galaxy, with the fragments dispersed in the ICM, 
the number of galaxies $N_{\rm accr}$ destroyed by tides
caused by a massive galaxy,
with the fragments being accreted onto that galaxy, the
number of galaxies $N_{\rm tides}^{\rm halo}$ destroyed by the tidal field 
of the background halo, and
the number of galaxies $N_{\rm eject}$ ejected from the cluster,
respectively. 
Column~9 shows the fraction {\it by numbers} of galaxies 
$f_{\rm surv}$ that survive to the present. 

We did not find a single occurrence of a galaxy destroyed
by tides from the background halo, 
and the number of galaxies ejected is either 0 or 1. 
There are large variations in the other numbers from one run to the next, 
but some trends are apparent. 
Typically, 50\% to 60\% of the galaxies are destroyed. 
Run~A2 is an extreme case, with 78\% of the galaxies being destroyed. 
The destruction of galaxies by mergers dominates over the destruction 
by tides, by more than a factor of 2 except for run A7. 
If we treat the cases of tidal disruption followed by accretion as being mergers, 
then mergers dominate even more over tidal disruption. 
When galaxies are destroyed by tides, the dispersion of fragments into the 
ICM always dominates over the accretion of fragments onto the massive galaxy, 
but the ratio varies wildly, from 114:7 in run~A7 to 66:51 in run~A8. 

We can now evaluate the fraction $f_{\rm ICS}^{\phantom1}$ 
of the total luminosity of the 
cluster that comes from intracluster stars. Since we assume a constant 
mass-to-light ratio, this fraction is given by
\begin{equation}
f_{\rm ICS}^{\phantom1}={M_{\rm tides}^{\rm gal}+M_{\rm tides}^{\rm halo}
\over M_{\rm total}-M_{\rm eject}}\,.
\end{equation}

\noindent where the letter $M$ refers to the mass in galaxies, rather than
their number. 
The galactic mass contribution to the ICM consists of galaxies destroyed by 
tides of another more massive galaxy, and by tides of the background halo
(though there are no such cases in this series).

The values of $f_{\rm ICS}^{\phantom1}$ are listed in column 10 of
Table~\ref{seriesa}.
Again, there are large variations. 
In particular, the fraction is very large for run~A2, and very small 
for runs~A3 and~A5. 
Averaging over all runs, we get
\begin{equation}
f_{\rm ICS}^{\phantom1}=0.254\pm0.093\,.
\end{equation} 

\noindent
Even though, in most cases about half the number of galaxies are 
destroyed, they tend
to be low-mass galaxies, which explains why $f_{\rm ICS}<1-f_{\rm surv}$,
for all the runs.

The galaxies being destroyed by mergers and tides, or escaping are mostly 
low-mass galaxies. This leads to a flattening of the galaxy mass 
distribution function. 
We computed the best numerical fit to the Schechter luminosity function 
exponent $\alpha$ [eq.~(\ref{schechter})] for the surviving galaxies at the end of 
the simulations. 
This is listed as $\alpha_{\rm end}^{\phantom1}$ in column~12 of 
Table~\ref{seriesa}. Averaging over all runs, we get
\begin{equation}
\alpha_{\rm end}^{\phantom1}=-1.206\pm0.040\,.
\end{equation}

\subsection{Series B: Turning on Harassment} 
\label{sec-seriesB}

\begin{deluxetable*}{cccccccccccc}
\tabletypesize{\scriptsize}
\tablecaption{Simulations for Series B}
\tablewidth{0pt}
\tablehead{
\colhead{Run} & \colhead{$M_{\rm total}[10^{11}M_\odot]$} 
& \colhead{$N_{\rm total}$} &
\colhead{$N_{\rm merge}$} & \colhead{$N_{\rm tides}^{\rm gal}$} & 
\colhead{$N_{\rm accr}$} & \colhead{$N_{\rm tides}^{\rm halo}$} & 
\colhead{$N_{\rm eject}$} &
\colhead{$f_{\rm surv}^{\phantom1}$} &
\colhead{$f_{\rm ICS}^{\phantom1}$} &
\colhead{$\alpha_{\rm start}^{\phantom1}$} & 
\colhead{$\alpha_{\rm end}^{\phantom1}$} 
}
\startdata
 B1 &  855.0 & 440 & 134 &  62 &  6 & 0 & 1 & 0.539 & 0.237 & $-1.26$ & $-1.24$ \cr
 B2 &  862.0 & 702 & 337 & 210 & 25 & 0 & 1 & 0.184 & 0.454 & $-1.28$ & $-1.11$ \cr
 B3 & 1100.4 & 480 & 126 &  48 &  7 & 0 & 0 & 0.623 & 0.139 & $-1.28$ & $-1.24$ \cr
 B4 &  831.0 & 530 & 159 &  94 & 12 & 0 & 0 & 0.500 & 0.184 & $-1.26$ & $-1.15$ \cr
 B5 &  980.4 & 459 & 171 &  78 &  7 & 0 & 1 & 0.440 & 0.154 & $-1.25$ & $-1.21$ \cr
 B6 &  967.8 & 640 & 226 &  85 &  6 & 0 & 0 & 0.505 & 0.173 & $-1.29$ & $-1.23$ \cr
 B7 &  720.0 & 579 & 249 & 134 &  9 & 0 & 0 & 0.323 & 0.364 & $-1.28$ & $-1.20$ \cr
 B8 &  757.8 & 435 & 172 &  79 & 62 & 0 & 1 & 0.278 & 0.271 & $-1.27$ & $-1.27$ \cr
 B9 &  925.6 & 514 & 223 & 100 & 12 & 0 & 1 & 0.346 & 0.270 & $-1.28$ & $-1.19$ \cr
B10 &  858.9 & 525 & 206 &  68 &  7 & 0 & 0 & 0.465 & 0.298 & $-1.31$ & $-1.25$ \cr
B11 &  880.2 & 547 & 233 &  85 &  9 & 0 & 1 & 0.400 & 0.280 & $-1.33$ & $-1.27$ \cr
B12 &  826.1 & 548 & 241 &  92 & 19 & 0 & 1 & 0.356 & 0.240 & $-1.29$ & $-1.20$ \cr
B13 & 1041.5 & 431 & 160 &  58 &  4 & 0 & 0 & 0.485 & 0.269 & $-1.27$ & $-1.26$ \cr
B14 &  957.3 & 486 & 173 &  59 &  5 & 0 & 0 & 0.512 & 0.274 & $-1.29$ & $-1.22$ \cr
B15 &  860.5 & 520 & 155 &  86 & 17 & 0 & 1 & 0.502 & 0.277 & $-1.26$ & $-1.19$ \cr
B16 &  858.0 & 483 & 264 &  90 &  6 & 0 & 0 & 0.255 & 0.354 & $-1.28$ & $-1.16$ \cr
B17 &  792.6 & 451 & 178 &  83 & 16 & 0 & 1 & 0.384 & 0.346 & $-1.24$ & $-1.13$ \cr
\enddata
\label{seriesb}
\end{deluxetable*}

We modified the algorithm to include the effect of galaxy harassment
(see \S\ref{sec-harass}), and rerun the calculations of series~A with
the same initial conditions. We also added one more run, B17. 
The results are shown in Table~\ref{seriesb}, 
which follows the same format as Table~\ref{seriesa}. 
Comparing with series~A, the number of galaxies destroyed
by mergers is very similar, but the number of galaxies destroyed by tides tends
to be significantly higher. For instance, it goes from 67 to 94 for runs~A4-B4,
and from 64 to 86 for runs A15-B15. 
This is because, when a galaxy is subjected to harassment, 
its binding energy is reduced, and it becomes more prone to experience 
tidal disruption later. The number of tidal disruptions followed by accretion
does not change significantly, though. Hence, the additional, tidally-disrupted
galaxies almost all contribute to the intracluster stars. 
The values of $f_{\rm ICS}$
are therefore increased relative to series~A. The mean value is
\begin{equation}
f_{\rm ICS}^{\phantom1}=0.269\pm0.081\,.
\end{equation}

\noindent
This is not significantly larger than for series~A. The additional galaxies
destroyed are mostly low-mass galaxies. 
We also recalculated the best-fit Schechter exponent $\alpha$ 
for the surviving galaxies at $z=0$. The mean value for the runs in 
this series is

\begin{equation}
\alpha_{\rm end}^{\phantom1}=-1.207\pm0.048\,.
\end{equation}

\begin{figure}
\begin{center}
\includegraphics[width=3.5in]{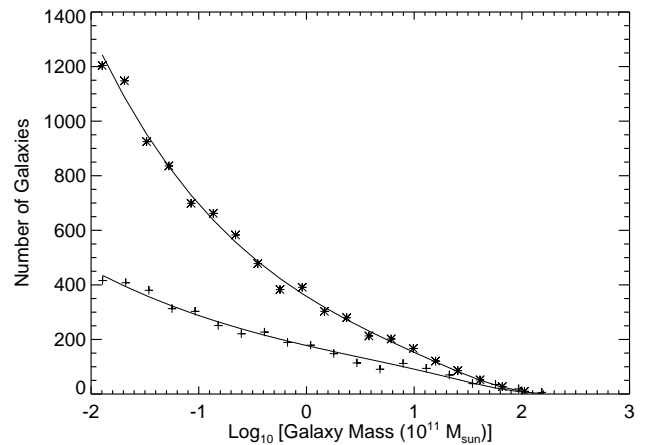}
\caption{Mass distribution function for galaxies in Series B,
obtained by adding the numbers for all runs. 
Results are plotted for initial 8770 galaxies at $z=1$ (asterisks) 
and surviving 3614 galaxies at $z=0$ (plus signs). 
The curves show the best fit of a Schechter distribution function 
(eq.~\ref{schechter}), 
with $\alpha=-1.28$ at $z=1$ (upper curve), 
and $\alpha=-1.20$ at $z=0$ (lower curve).}
\label{Schechter_B}
\end{center}
\end{figure}

Figure~\ref{Schechter_B} shows the total galaxy counts in mass bins, 
obtained by combining all the runs in series~B, 
along with the fitting curves to a Schechter distribution function 
[eq.~(\ref{schechter})]. 
The best fit Schechter exponent for the initial galaxy distribution 
(the upper curve at $z=1$) is $\alpha = -1.28$, and for the 
final surviving galaxy distribution (the lower curve at $z=0$) 
is $\alpha = -1.20$. 
These values of $\alpha$ were obtained by performing the 
numerical Schechter function fits on the combined set of galaxies 
taken from all the 17 runs in this series, 
which amounts to 8770 initial galaxies at $z=1$ and 3614 surviving 
galaxies at $z=0$. 

Clearly from Figure~\ref{Schechter_B}, the fit at $z=0$ 
(lower curve) is excellent. 
This shows that, in our simulations, a Schechter mass (luminosity) distribution 
function at $z=1$ remains a Schechter distribution all the way to $z=0$, 
though half of the galaxies are destroyed. 
Only the slope $\alpha$ changes with time. 

\subsection{Series C: Steeping up the Mass Distribution Function}
\label{sec-seriesC}

In the simulations of series~A and B, the Schechter exponent $\alpha$ 
evolves from
$\alpha\simeq-1.28$ at $z=1$ to $\alpha\simeq-1.21$ at $z=0$. 
Analyzing the combined set of galaxies of all the 17 runs in series~B, 
we obtained the best fit Schechter exponent 
for the surviving galaxy distribution at $z=0$ as $\alpha = -1.20$. 
This is a problem, since the value of $\alpha= -1.28$ 
is based on observations of nearby clusters \citep{deproprisetal2003}, 
and should be valid for clusters of galaxies at $z=0$,
whereas we used the same $\alpha=-1.28$ to start our simulations at $z=1$, 
and it flattened to $\alpha = -1.20$ at $z=0$. 
\citet{ryanetal07} recently determined the luminosity function
of a large sample of galaxies at $z\simeq1$, and concluded that
there is a steepening of the faint-end slope with redshift, 
which is expected in the hierarchical formation scenario of galaxies. 
They obtained a value of the faint-end slope $\alpha=-1.32\pm0.07$ at $z=1$. 

In our simulations we take into account this flattening of the luminosity function 
over time as explained below. 
The results of series~A and B suggest that $|\alpha|$ decreases by $\sim0.08$ 
between $z=1$ and $z=0$. Using this as a guide, 
we performed a new series of simulations,
series~C, using $\alpha_{\rm start}=-1.36$, with the hope that this value
will evolve toward something close to $\alpha=-1.28$ at $z=0$. 
The results are shown in Table~\ref{seriesc}. 
The average values of $\alpha$ are
\begin{eqnarray}
\alpha_{\rm start}^{\phantom1}&=&-1.357\pm0.021\,,\\
\alpha_{\rm end}^{\phantom1}&=&-1.272\pm0.050\,.
\end{eqnarray}

\begin{deluxetable*}{cccccccccccc}
\tabletypesize{\scriptsize}
\tablecaption{Simulations for Series C}
\tablewidth{0pt}
\tablehead{
\colhead{Run} & \colhead{$M_{\rm total}[10^{11}M_\odot]$} 
& \colhead{$N_{\rm total}$} &
\colhead{$N_{\rm merge}$} & \colhead{$N_{\rm tides}^{\rm gal}$} & 
\colhead{$N_{\rm accr}$} & \colhead{$N_{\rm tides}^{\rm halo}$} & 
\colhead{$N_{\rm eject}$} &
\colhead{$f_{\rm surv}^{\phantom1}$} &
\colhead{$f_{\rm ICS}^{\phantom1}$} &
\colhead{$\alpha_{\rm start}^{\phantom1}$} & 
\colhead{$\alpha_{\rm end}^{\phantom1}$} 
}
\startdata
 C1 & 1144.3 &  948 & 322 & 123 & 13 & 0 & 7 & 0.509 & 0.162 & $-1.34$ & $-1.27$ \cr
 C2 & 1009.9 &  936 & 283 & 138 & 19 & 0 & 1 & 0.529 & 0.195 & $-1.38$ & $-1.30$ \cr
 C3 &  914.2 &  603 & 316 &  97 & 13 & 0 & 0 & 0.294 & 0.265 & $-1.35$ & $-1.24$ \cr
 C4 &  970.5 &  635 & 162 &  86 & 11 & 0 & 1 & 0.591 & 0.216 & $-1.32$ & $-1.28$ \cr
 C5 &  935.9 &  672 & 250 & 101 &  7 & 0 & 1 & 0.466 & 0.158 & $-1.33$ & $-1.25$ \cr
 C6 &  948.2 & 1136 & 502 & 268 &  6 & 0 & 0 & 0.317 & 0.318 & $-1.35$ & $-1.21$ \cr
 C7 &  926.4 &  884 & 276 & 155 & 10 & 0 & 1 & 0.500 & 0.258 & $-1.35$ & $-1.28$ \cr
 C8 &  818.4 &  696 & 347 & 142 & 62 & 0 & 1 & 0.207 & 0.423 & $-1.35$ & $-1.26$ \cr
 C9 &  875.9 &  903 & 421 & 255 & 44 & 0 & 1 & 0.202 & 0.405 & $-1.34$ & $-1.19$ \cr
C10 & 1110.5 &  785 & 286 & 100 & 20 & 0 & 1 & 0.482 & 0.168 & $-1.39$ & $-1.39$ \cr
C11 &  864.2 &  809 & 338 & 139 & 18 & 0 & 1 & 0.387 & 0.386 & $-1.40$ & $-1.36$ \cr
C12 & 1053.2 &  978 & 395 & 212 & 29 & 0 & 0 & 0.350 & 0.294 & $-1.36$ & $-1.29$ \cr
C13 & 1151.7 &  790 & 344 & 127 & 24 & 0 & 1 & 0.372 & 0.262 & $-1.37$ & $-1.29$ \cr
C14 &  876.8 &  684 & 335 & 133 & 23 & 0 & 1 & 0.281 & 0.320 & $-1.37$ & $-1.21$ \cr
C15 &  788.0 &  769 & 240 & 140 & 27 & 0 & 0 & 0.471 & 0.243 & $-1.35$ & $-1.26$ \cr
C16 &  905.1 &  578 & 284 &  80 & 24 & 0 & 1 & 0.327 & 0.430 & $-1.36$ & $-1.29$ \cr
C17 &  889.2 &  822 & 325 & 169 & 13 & 0 & 0 & 0.383 & 0.323 & $-1.36$ & $-1.25$ \cr
\enddata
\label{seriesc}
\end{deluxetable*}

\begin{figure}
\begin{center}
\includegraphics[width=3.5in]{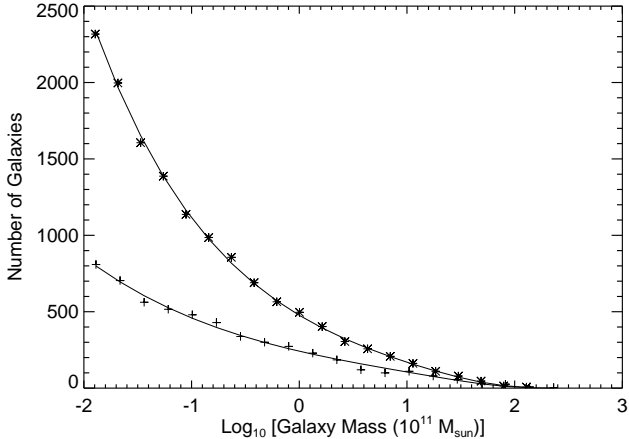}
\caption{Same as Figure~\ref{Schechter_B}, for series C. 
Results are plotted for initial 13628 galaxies at $z=1$ (asterisks) 
and surviving 5356 galaxies at $z=0$ (plus signs). 
The best fit Schechter distribution functions are 
with $\alpha=-1.36$ at $z=1$ (upper curve), 
and $\alpha=-1.27$ at $z=0$ (lower curve).}
\label{Schechter_C}
\end{center}
\end{figure}

Figure~\ref{Schechter_C} (analogous to Figure~\ref{Schechter_B}) 
shows that at $z=0$, a Schechter distribution function is still a 
good approximation to the mass distribution. 
Combining the galaxies from all the 17 runs in this series 
(13628 initial galaxies at $z=1$, and 5356 surviving galaxies at $z=0$), 
the best fit Schechter exponent for the initial galaxy distribution 
(the upper curve) is $\alpha = -1.36$, and for the 
surviving galaxy distribution (the lower curve) is $\alpha = -1.27$. 
This value of $\alpha$ is close enough to our target value
of $-1.28$. So from now on, in all subsequent series with the $\beta$ 
model halo density profile, we will use an initial $\alpha$ of $-1.36$, 
as shown in Table~\ref{series}. This value is well inside the range obtained
by \citet{ryanetal07}.

Using a steeper galaxy distribution while still requiring that the clusters contain 
25 galaxies with $L>0.2L_*$ (see initial conditions in \S\ref{sec-initial}) 
results in the initial number of galaxies 
being larger by a factor of about 2 (column 3 of Table~\ref{seriesc}). 
But the numbers of galaxies destroyed by mergers and 
tides are also higher relative to series~B. As a result the trends are similar. 
In particular, mergers still dominate over tides by more than a factor of 2. 

The run C1 has a larger number of galaxies ejected from the cluster. 
This is because the most massive galaxy, located at the center of the cluster,
was particularly large. Its radius was $s=385\,\rm kpc$, compared to
$s\lesssim300\,\rm kpc$ for the other runs. 
This increased the value of $R_0$ (see \S\ref{sec-initial}) used for 
setting up the initial conditions. As a result, more galaxies were 
located at larger radii, where they are more likely to escape.

The mean value of $f_{\rm ICS}$ for this series is
\begin{equation}
f_{\rm ICS}^{\phantom1}=0.284\pm0.090\,.
\end{equation}

\subsection{Series D: Adding a cD Galaxy} 
\label{sec-seriesD}

A cD ({\it central dominant}) galaxy is a very bright supergiant 
elliptical galaxy with an extended envelope (or a {\it diffuse} halo) 
found at the center of a cluster \citep{schombert88}. 
Several galaxy clusters have been found to have cD galaxies at their centers 
(e.g., \citealt{quintana82, hill98, oegerle01, jordan04, seigar07}). 
We performed some simulations by incorporating a cD galaxy in the clusters. 
Being the brightest and most massive cluster galaxy, the mass of a cD is 
larger than 
the prediction of the normal Schechter distribution function (eq.~\ref{schechter}). 
So we introduced the cD galaxy manually at our simulated cluster center. 
We adopted a luminosity of $L_{\rm cD} = 10L^*$, 
which is a canonical value for a cD. 
Using our mass to light ratio ($\Upsilon=193~h~M_\odot/L_\odot$, 
\S\ref{sec-initial}), 
this corresponds to a cD galaxy mass of $M_{\rm cD} = 437.6 \ee{11} \msun$. 
When we wanted a cD galaxy present in the simulation 
we changed the mass of the cluster central galaxy (see \S\ref{sec-initial}) 
to the cD mass, $M_{\rm cD}$. 
This allowed us to keep the appropriate initial galaxy distribution for a cluster 
while incorporating a cD galaxy at rest, located at the center. 

\begin{deluxetable*}{cccccccccccc}
\tabletypesize{\scriptsize}
\tablecaption{Simulations for Series D}
\tablewidth{0pt}
\tablehead{
\colhead{Run} & \colhead{$M_{\rm total}[10^{11}M_\odot]$} 
& \colhead{$N_{\rm total}$} &
\colhead{$N_{\rm merge}$} & \colhead{$N_{\rm tides}^{\rm gal}$} & 
\colhead{$N_{\rm accr}$} & \colhead{$N_{\rm tides}^{\rm halo}$} & 
\colhead{$N_{\rm eject}$} &
\colhead{$f_{\rm surv}^{\phantom1}$} &
\colhead{$f_{\rm ICS}^{\phantom1}$} &
\colhead{$\alpha_{\rm start}^{\phantom1}$} & 
\colhead{$\alpha_{\rm end}^{\phantom1}$} 
}
\startdata
 D1 & 1394.6 & 1023 & 345 & 165 & 83 & 0 & 0 & 0.420 & 0.127 & $-1.38$ & $-1.33$ \cr
 D2 & 1374.4 & 866 & 330 &  97 &  37 & 0 & 1 & 0.463 & 0.085 & $-1.38$ & $-1.33$ \cr
 D3 & 1431.1 & 906 & 257 & 199 & 187 & 0 & 0 & 0.290 & 0.125 & $-1.39$ & $-1.32$ \cr
 D4 & 1163.2 & 754 & 211 & 152 & 120 & 0 & 0 & 0.359 & 0.117 & $-1.33$ & $-1.26$ \cr
 D5 & 1239.5 & 788 & 283 & 167 & 197 & 0 & 0 & 0.179 & 0.167 & $-1.33$ & $-1.21$ \cr
 D6 & 1158.8 & 527 & 131 &  59 &  53 & 0 & 1 & 0.537 & 0.065 & $-1.34$ & $-1.30$ \cr
 D7 & 1300.7 & 663 & 236 &  85 &  73 & 0 & 1 & 0.404 & 0.151 & $-1.37$ & $-1.31$ \cr
 D8 & 1298.4 & 773 & 296 & 113 & 116 & 0 & 0 & 0.321 & 0.120 & $-1.36$ & $-1.33$ \cr
 D9 & 1299.5 & 652 & 256 &  95 & 63 & 0 & 1 & 0.364 & 0.141 & $-1.39$ & $-1.33$ \cr
D10 & 1169.4 & 589 & 193 &  76 & 67 & 0 & 1 & 0.428 & 0.106 & $-1.37$ & $-1.32$ \cr
D11 & 1145.8 & 741 & 282 & 143 & 79 & 0 & 1 & 0.318 & 0.124 & $-1.38$ & $-1.25$ \cr
D12 & 1227.5 & 732 & 235 &  57 & 22 & 0 & 9 & 0.559 & 0.075 & $-1.39$ & $-1.35$ \cr
D13 & 1334.2 & 952 & 330 & 166 & 70 & 0 & 0 & 0.406 & 0.151 & $-1.35$ & $-1.31$ \cr
D14 & 1286.0 & 844 & 283 & 184 & 115 & 0 & 1 & 0.309 & 0.203 & $-1.34$ & $-1.24$ \cr
D15 & 1470.8 & 888 & 298 & 115 & 56 & 0 & 0 & 0.472 & 0.140 & $-1.36$ & $-1.33$ \cr
D16 & 1454.4 & 966 & 288 & 170 & 125 & 0 & 0 & 0.396 & 0.086 & $-1.34$ & $-1.30$ \cr
\enddata
\label{seriesd}
\end{deluxetable*}

We performed simulations by adding a cD galaxy to our 
Virgo-like cluster, and called it series~D. 
The results are listed in Table~\ref{seriesd}, 
from which certain trends are clear after incorporating a cD galaxy 
in the simulation. 
The total galaxy mass increases since a massive cD galaxy is being added. 
More prominent than in the previous series~A, B, and C, 
here galaxy mergers outnumber tides by factors $\sim 2-3$, 
which go as high as 4 in run D12. 

A striking new feature in cases incorporating a cD galaxy is the increase in the number of 
accretions after tidal disruption by a galaxy, fully 1/4 of the galaxies being acreted in run D5. 
Since in these accretions, the smaller galaxy is tidally destroyed 
and is absorbed (or merged) by the massive galaxy (\S\ref{sec-tideAccr}), it appears, in our simulated clusters, that in the presence of a cD galaxy
the number of effective mergers is very high. 

The mass fraction imparted to ICS has decreased in all the runs, 
with a value as low as 0.065 in run D6. 
To explain such a result, we note that the most massive central galaxy 
(cD or otherwise) in our simulated cluster is never destroyed 
because of its large mass. 
In an encounter, it is normally the lower-mass galaxy that gets destroyed. 
Also the initial conditions of the most massive galaxy 
(at rest at the center, see \S\ref{sec-initial}) 
make it less likely to be destroyed by the tidal field of the halo. 
If the central galaxy is a cD, a large mass fraction (as high as 38\% in run D11) 
is locked into it, which can never contribute to the ICS. 
So a smaller mass fraction is available to be transferred to the ICS, 
which eventually leads to a decrease in $f_{\rm ICS}$. 

The mean value of $f_{\rm ICS}$ for series~D is
\begin{equation}
f_{\rm ICS}^{\phantom1}=0.124\pm0.036\,.
\end{equation}

\subsection{Series E \& F: Other $\beta$ Profiles}
\label{sec-seriesEF}

\begin{deluxetable*}{cccccccccccc}
\tabletypesize{\scriptsize}
\tablecaption{Simulations for Series E}
\tablewidth{0pt}
\tablehead{
\colhead{Run} & \colhead{$M_{\rm total}[10^{11}M_\odot]$} 
& \colhead{$N_{\rm total}$} &
\colhead{$N_{\rm merge}$} & \colhead{$N_{\rm tides}^{\rm gal}$} & 
\colhead{$N_{\rm accr}$} & \colhead{$N_{\rm tides}^{\rm halo}$} & 
\colhead{$N_{\rm eject}$} &
\colhead{$f_{\rm surv}^{\phantom1}$} &
\colhead{$f_{\rm ICS}^{\phantom1}$} &
\colhead{$\alpha_{\rm start}^{\phantom1}$} & 
\colhead{$\alpha_{\rm end}^{\phantom1}$} 
}
\startdata
 E1 & 1065.4 & 1023 & 406 & 159 & 24 &  8 &  1 & 0.415 & 0.297 & $-1.38$ & $-1.31$ \cr
 E2 & 1034.0 &  866 & 326 & 104 &  7 &  6 &  2 & 0.486 & 0.168 & $-1.38$ & $-1.31$ \cr
 E3 & 1051.3 &  906 & 405 & 213 & 19 & 14 &  0 & 0.282 & 0.446 & $-1.38$ & $-1.27$ \cr
 E4 &  799.4 &  754 & 250 & 177 & 35 &  7 &  0 & 0.378 & 0.391 & $-1.34$ & $-1.25$ \cr
 E5 &  868.7 &  788 & 394 & 201 & 42 & 12 &  0 & 0.176 & 0.492 & $-1.32$ & $-1.22$ \cr
 E6 &  795.8 &  527 & 154 &  67 &  7 &  6 &  1 & 0.554 & 0.303 & $-1.34$ & $-1.29$ \cr
 E7 &  947.7 &  663 & 276 &  97 &  9 &  3 &  1 & 0.418 & 0.376 & $-1.37$ & $-1.31$ \cr
 E8 &  941.0 &  773 & 395 & 121 & 11 &  8 &  0 & 0.307 & 0.348 & $-1.37$ & $-1.28$ \cr
 E9 &  934.5 &  652 & 263 & 109 & 10 &  6 &  1 & 0.403 & 0.325 & $-1.39$ & $-1.34$ \cr
E10 &  807.5 &  589 & 210 &  98 & 11 &  5 &  1 & 0.325 & 0.448 & $-1.38$ & $-1.35$ \cr
E11 &  760.7 &  741 & 349 & 142 &  9 &  8 &  1 & 0.313 & 0.283 & $-1.48$ & $-1.23$ \cr
E12 &  946.4 &  732 & 227 &  57 &  4 &  2 & 28 & 0.566 & 0.149 & $-1.39$ & $-1.35$ \cr
E13 &  953.4 &  952 & 344 & 180 & 14 & 11 &  0 & 0.423 & 0.349 & $-1.35$ & $-1.29$ \cr
E14 &  950.5 &  844 & 325 & 197 & 23 &  6 &  1 & 0.346 & 0.395 & $-1.33$ & $-1.29$ \cr
E15 & 1145.5 &  888 & 320 & 118 & 13 &  5 &  0 & 0.486 & 0.339 & $-1.36$ & $-1.29$ \cr
E16 & 1125.4 &  966 & 439 & 146 & 10 & 10 &  1 & 0.373 & 0.393 & $-1.34$ & $-1.26$ \cr
\enddata
\label{seriese}
\end{deluxetable*}

In the next two series of runs, we consider a different background halo. We use
a $\beta$-profile with $\beta = 0.53$, a core radius $r_c=28\,\rm kpc$ and
a central density $\rho_0=7.27\times10^{-26}\,\rm g\,cm^{-3}$, 
which is appropriate for a cluster like Perseus \citep{piffarettietal06}.
Series~E and F do not include, and include a cD galaxy, respectively
(hence series~E should be compared with series~C, and series~F with series~D). 

The results for series~E are shown in Table~\ref{seriese}. The most notable feature is
that some galaxies are destroyed by the tidal field of the background halo, 
something that never happened with Virgo-like clusters. 
In order to explain such a behavior we note that tidal disruption by the 
cluster halo generally 
occurs with galaxies at a distance $r \lesssim 1$ Mpc from the cluster center. 
Our simulated Perseus-like cluster halo 
mass profile rises more steeply than the Virgo-like cluster up to $\sim 1.7$ Mpc, 
making Perseus more massive in the inner regions. 
So a galaxy at a smaller distance, precisely at $r \lesssim 1.7$ Mpc, 
from the cluster center feels a larger tidal field from a more massive halo, 
and is more prone to be disrupted in the Perseus-like cluster.
                                                           
The numbers of other galaxy outcomes are comparable for Perseus-like and 
Virgo-like clusters,
with mergers dominating over tides. 
The mean $f_{\rm ICS}$ for series~E is
\begin{equation}
f_{\rm ICS}^{\phantom1}=0.344\pm0.093\,.
\end{equation}

\noindent This $f_{\rm ICS}$ is somewhat larger than the 
Virgo-like cluster mean (series~C). 
This can be attributed to the non-zero tidal disruption by the cluster halo, resulting here 
in a finite contribution to the ICS mass fraction. 

\begin{deluxetable*}{cccccccccccc}
\tabletypesize{\scriptsize}
\tablecaption{Simulations for Series F}
\tablewidth{0pt}
\tablehead{
\colhead{Run} & \colhead{$M_{\rm total}[10^{11}M_\odot]$} 
& \colhead{$N_{\rm total}$} &
\colhead{$N_{\rm merge}$} & \colhead{$N_{\rm tides}^{\rm gal}$} & 
\colhead{$N_{\rm accr}$} & \colhead{$N_{\rm tides}^{\rm halo}$} & 
\colhead{$N_{\rm eject}$} &
\colhead{$f_{\rm surv}^{\phantom1}$} &
\colhead{$f_{\rm ICS}^{\phantom1}$} &
\colhead{$\alpha_{\rm start}^{\phantom1}$} & 
\colhead{$\alpha_{\rm end}^{\phantom1}$} 
}
\startdata
 F1 & 1394.6 & 1023 & 386 & 153 & 88 & 4 & 1 & 0.382 & 0.138 & $-1.38$ & $-1.28$ \cr
 F2 & 1374.4 & 866 & 304 & 96 & 47 & 1 & 2 & 0.480 & 0.106 & $-1.38$ & $-1.33$ \cr
 F3 & 1431.1 & 906 & 277 & 183 & 196 & 1 & 0 & 0.275 & 0.144 & $-1.39$ & $-1.25$ \cr
 F4 & 1163.2 & 754 & 211 & 180 & 109 & 0 & 0 & 0.337 & 0.126 & $-1.33$ & $-1.27$ \cr
 F5 & 1239.5 & 788 & 288 & 166 & 199 & 0 & 0 & 0.171 & 0.158 & $-1.33$ & $-1.21$ \cr
 F6 & 1158.8 & 527 & 141 & 60 & 55 & 1 & 1 & 0.510 & 0.092 & $-1.34$ & $-1.29$ \cr
 F7 & 1300.7 & 663 & 253 & 81 & 70 & 1 & 1 & 0.388 & 0.113 & $-1.37$ & $-1.28$ \cr
 F8 & 1298.4 & 773 & 295 & 124 & 108 & 2 & 0 & 0.316 & 0.134 & $-1.36$ & $-1.29$ \cr
 F9 & 1299.5 & 652 & 226 & 93 & 61 & 2 & 1 & 0.413 & 0.153 & $-1.39$ & $-1.38$ \cr
F10 & 1169.4 & 589 & 182 & 83 & 72 & 0 & 1 & 0.426 & 0.133 & $-1.37$ & $-1.33$ \cr
F11 & 1145.8 & 741 & 269 & 148 & 76 & 1 & 1 & 0.332 & 0.154 & $-1.38$ & $-1.29$ \cr
F12 & 1227.5 & 732 & 194 & 58 & 26 & 0 & 31 & 0.578 & 0.067 & $-1.39$ & $-1.36$ \cr
F13 & 1334.1 & 952 & 322 & 175 & 85 & 1 & 0 & 0.388 & 0.121 & $-1.35$ & $-1.29$ \cr
F14 & 1286.0 & 844 & 272 & 180 & 107 & 0 & 1 & 0.336 & 0.197 & $-1.34$ & $-1.30$ \cr
F15 & 1470.8 & 888 & 306 & 103 & 54 & 1 & 0 & 0.478 & 0.108 & $-1.36$ & $-1.31$ \cr
F16 & 1454.4 & 966 & 314 & 155 & 135 & 2 & 1 & 0.372 & 0.097 & $-1.34$ & $-1.29$ \cr
\enddata
\label{seriesf}
\end{deluxetable*}

Table~\ref{seriesf} shows the results for series~F, i.e., 
simulations of a Perseus-like cluster with a cD galaxy at the center. 
Here few galaxies are destroyed by the tidal field of the cluster halo, 
yet the number is smaller than in series E.  It appears, then that the presence of a cD galaxy reduces the number of tidal disruptions 
by the background halo, since galaxies that would be destroyed by the 
tidal field of the central parts of the halo 
are being destroyed by the cD galaxy instead.

Comparing the results for series~D and series~F (Virgo-like 
and Perseus-like clusters with a cD galaxy,)
the numbers-- merger, galaxy-tide and accretion are similar. 
Series~F continues the trend of increased accretions when a cD galaxy is introduced. 
Also series~F has a smaller fraction of mass going to ICS. 
The mean $f_{\rm ICS}$ for series~F is
\begin{equation}
f_{\rm ICS}^{\phantom1}=0.128\pm0.031\,.
\end{equation}

\noindent This $f_{\rm ICS}$ is very similar to that of the relevant 
Virgo-like cluster mean (series~D). 
This implies that in the presence of a cD galaxy, the ICS mass fraction is not 
so sensitive to the parameters of the $\beta$-model density profile. 

\subsection{Series G, H, \& I: NFW Profile}
\label{sec-seriesGHI} 

We now consider a background halo described by a NFW profile 
[see \S\ref{sec-haloRho}, eq.~(\ref{rho-anal})], 
with a scale radius $r_s = 200$ kpc, and a concentration parameter $c = 5$. 
These values are adopted from observational studies of galaxy clusters 
\citep{arabadjis02, pratt05, maughan07} where the authors found the best fitting 
NFW model parameters for cluster mass profiles. 

We do not necessarily expect the flattening of the Schechter mass
function to be the same for the NFW profile halo and the $\beta$-profile
halo. 
So at first we performed a series with $\alpha=-1.28$ (see Table~\ref{series}), 
and called it series~G. 
The results are listed in Table~\ref{seriesg}. 

\begin{deluxetable*}{cccccccccccc}
\tabletypesize{\scriptsize}
\tablecaption{Simulations for Series G}
\tablewidth{0pt}
\tablehead{
\colhead{Run} & \colhead{$M_{\rm total}[10^{11}M_\odot]$} 
& \colhead{$N_{\rm total}$} &
\colhead{$N_{\rm merge}$} & \colhead{$N_{\rm tides}^{\rm gal}$} & 
\colhead{$N_{\rm accr}$} & \colhead{$N_{\rm tides}^{\rm halo}$} & 
\colhead{$N_{\rm eject}$} &
\colhead{$f_{\rm surv}^{\phantom1}$} &
\colhead{$f_{\rm ICS}^{\phantom1}$} &
\colhead{$\alpha_{\rm start}^{\phantom1}$} & 
\colhead{$\alpha_{\rm end}^{\phantom1}$} 
}
\startdata
 G1 & 721.9 & 372 & 72 & 55 & 6 & 47 & 1 & 0.513 & 0.486 & $-1.27$ & $-1.26$ \cr
 G2 & 1034.3 & 637 & 129 & 80 & 4 & 63 & 7 & 0.556 & 0.405 & $-1.29$ & $-1.26$ \cr 
 G3 & 821.1 & 530 & 127 & 89 & 5 & 73 & 1 & 0.443 & 0.587 & $-1.31$ & $-1.29$ \cr
 G4 & 992.3 & 457 & 113 & 61 & 9 & 44 & 1 & 0.501 & 0.211 & $-1.28$ & $-1.26$ \cr
 G5 & 899.7 & 618 & 94 & 56 & 7 & 64 & 8 & 0.629 & 0.443 & $-1.29$ & $-1.28$ \cr
 G6 & 947.1 & 452 & 95 & 31 & 2 & 34 & 17 & 0.604 & 0.263 & $-1.31$ & $-1.32$ \cr
 G7 & 865.2 & 542 & 170 & 91 & 7 & 77 & 1 & 0.362 & 0.543 & $-1.31$ & $-1.25$ \cr
 G8 & 1011.6 & 725 & 169 & 101 & 7 & 71 & 1 & 0.519 & 0.502 & $-1.24$ & $-1.20$ \cr
 G9 & 1100.6 & 726 & 214 & 144 & 17 & 103 & 1 & 0.340 & 0.467 & $-1.27$ & $-1.23$ \cr 
G10 & 1174.4 & 619 & 190 & 76 & 2 & 57 & 6 & 0.465 & 0.458 & $-1.28$ & $-1.23$ \cr
\enddata
\label{seriesg}
\end{deluxetable*}

\begin{deluxetable*}{cccccccccccc}
\tabletypesize{\scriptsize}
\tablecaption{Simulations for Series H}
\tablewidth{0pt}
\tablehead{
\colhead{Run} & \colhead{$M_{\rm total}[10^{11}M_\odot]$} 
& \colhead{$N_{\rm total}$} &
\colhead{$N_{\rm merge}$} & \colhead{$N_{\rm tides}^{\rm gal}$} & 
\colhead{$N_{\rm accr}$} & \colhead{$N_{\rm tides}^{\rm halo}$} & 
\colhead{$N_{\rm eject}$} &
\colhead{$f_{\rm surv}^{\phantom1}$} &
\colhead{$f_{\rm ICS}^{\phantom1}$} &
\colhead{$\alpha_{\rm start}^{\phantom1}$} & 
\colhead{$\alpha_{\rm end}^{\phantom1}$} 
}
\startdata
 H1 & 894.3 & 755 & 171 & 152 & 5 & 122 & 0 & 0.404 & 0.620 & $-1.33$ & $-1.33$ \cr
 H2 & 866.2 & 556 & 195 & 68 & 5 & 92 & 1 & 0.351 & 0.563 & $-1.28$ & $-1.24$ \cr 
 H3 & 955.0 & 621 & 161 & 84 & 2 & 67 & 1 & 0.493 & 0.396 & $-1.33$ & $-1.35$ \cr
 H4 & 758.6 & 588 & 153 & 113 & 7 & 109 & 1 & 0.349 & 0.702 & $-1.30$ & $-1.30$ \cr
 H5 & 1103.6 & 997 & 197 & 132 & 5 & 68 & 22 & 0.575 & 0.375 & $-1.30$ & $-1.26$ \cr
 H6 & 1008.7 & 774 & 266 & 162 & 10 & 148 & 0 & 0.243 & 0.682 & $-1.29$ & $-1.24$ \cr
 H7 & 1027.4 & 639 & 101 & 40 & 4 & 47 & 28 & 0.656 & 0.316 & $-1.32$ & $-1.30$ \cr 
 H8 & 833.8 & 601 & 126 & 75 & 5 & 70 & 7 & 0.529 & 0.496 & $-1.29$ & $-1.28$ \cr
 H9 & 740.3 & 494 & 109 & 53 & 9 & 54 & 4 & 0.536 & 0.398 & $-1.30$ & $-1.28$ \cr 
H10 & 952.9 & 744 & 159 & 96 & 13 & 92 & 1 & 0.515 & 0.451 & $-1.35$ & $-1.32$ \cr 
H11 & 779.4 & 519 & 113 & 54 & 9 & 59 & 2 & 0.543 & 0.375 & $-1.31$ & $-1.25$ \cr
H12 & 993.6 & 741 & 163 & 86 & 3 & 57 & 10 & 0.570 & 0.306 & $-1.33$ & $-1.29$ \cr
H13 & 931.8 & 534 & 131 & 60 & 5 & 62 & 2 & 0.513 & 0.412 & $-1.31$ & $-1.33$ \cr
H14 & 1012.8 & 748 & 202 & 117 & 8 & 95 & 1 & 0.434 & 0.508 & $-1.30$ & $-1.26$ \cr 
\enddata
\label{seriesh}
\end{deluxetable*}

\begin{deluxetable*}{cccccccccccc}
\tabletypesize{\scriptsize}
\tablecaption{Simulations for Series I}
\tablewidth{0pt}
\tablehead{
\colhead{Run} & \colhead{$M_{\rm total}[10^{11}M_\odot]$} 
& \colhead{$N_{\rm total}$} &
\colhead{$N_{\rm merge}$} & \colhead{$N_{\rm tides}^{\rm gal}$} & 
\colhead{$N_{\rm accr}$} & \colhead{$N_{\rm tides}^{\rm halo}$} & 
\colhead{$N_{\rm eject}$} &
\colhead{$f_{\rm surv}^{\phantom1}$} &
\colhead{$f_{\rm ICS}^{\phantom1}$} &
\colhead{$\alpha_{\rm start}^{\phantom1}$} & 
\colhead{$\alpha_{\rm end}^{\phantom1}$} 
}
\startdata
I1 & 1189.1 & 495 & 114 &  53 & 26 &  75 & 1 & 0.457 & 0.322 & $-1.34$ & $-1.31$ \cr
I2 & 1255.5 & 705 & 183 & 100 & 16 &  87 & 0 & 0.452 & 0.326 & $-1.34$ & $-1.32$ \cr
I3 & 1189.9 & 643 & 152 & 127 & 35 & 119 & 0 & 0.327 & 0.391 & $-1.29$ & $-1.31$ \cr
I4 & 1153.5 & 548 & 108 &  78 &  8 &  66 & 1 & 0.524 & 0.255 & $-1.32$ & $-1.30$ \cr
I5 & 1309.4 & 821 & 228 & 139 & 19 &  87 & 1 & 0.423 & 0.341 & $-1.33$ & $-1.31$ \cr
I6 & 1330.8 & 596 & 105 &  56 &  2 &  54 & 9 & 0.621 & 0.265 & $-1.31$ & $-1.30$ \cr
I7 & 1315.8 & 775 & 215 & 126 & 37 & 105 & 0 & 0.377 & 0.299 & $-1.33$ & $-1.31$ \cr
I8 & 1188.1 & 691 & 118 &  91 & 12 &  71 & 1 & 0.576 & 0.248 & $-1.36$ & $-1.35$ \cr
I9 & 1236.6 & 574 & 103 &  73 & 12 &  56 & 1 & 0.573 & 0.295 & $-1.32$ & $-1.30$ \cr
I10& 1093.4 & 634 &  96 &  96 &  8 &  64 & 1 & 0.582 & 0.274 & $-1.31$ & $-1.29$ \cr
\enddata 
\label{seriesi} 
\end{deluxetable*}

To contrast a NFW-model cluster with a $\beta$-model cluster, 
Series~G should be compared with Series~B, since these are with $\alpha=-1.28$, 
include galaxy harassment, and no cD galaxy. 
The most striking feature is the large number of galaxies destroyed 
by the tidal field of the NFW cluster halo. 
This halo tidal disruption was nil (in the Virgo-like cluster) 
to a handful (in the Perseus-like cluster) for the $\beta$-model background halo. 
With the NFW profile, the number of halo tides is comparable to the galaxy tides, 
even exceeding the latter in runs G5 and G6. 

The reason for such a behavior is that 
the NFW halo mass profile rises much more steeply than the 
$\beta$-model mass profile of a Virgo-like cluster up to a distance $r \sim 1.9$ Mpc. 
So the NFW halo is significantly more massive (by factors as high as $4 -5$) 
than the $\beta$-model halo at distances $r \lesssim 1$ Mpc, 
where halo tides are dominant (as discussed in \S\ref{sec-seriesEF}). 
Consequently galaxies near the cluster center experience a
larger tidal field and are more likely to be tidally disrupted. 

This larger number of halo tides alters several results in our 
simulated NFW model cluster as compared to the $\beta$-model. 
The mergers exceed the galaxy tides, usually by factors 1.3-1.8 
(except runs G6 and G10, where the factors are 3 and 2.5). 
But when tides by galaxy and cluster halo are added together, 
they become comparable to or even exceed the number of mergers. 
The accretions are always small in number, 
and when added to mergers do not have much effect on the above. 

The mean $f_{\rm ICS}$ for series~G is
\begin{equation}
f_{\rm ICS}^{\phantom1}=0.436\pm0.117\,.
\end{equation} 

\noindent This is significantly larger than the ICS mass fraction 
obtained with the $\beta$-model clusters  The reason is again the numerous halo tides. 
Some massive galaxies are being destroyed by the tidal field of 
the NFW halo, when they come near the cluster center, and
this is contributing a large mass fraction to the ICS. 

In this series~G, we obtained the average values of $\alpha$ as 
\begin{eqnarray}
\alpha_{\rm start}^{\phantom1}&=&-1.285\pm0.022\,,\\
\alpha_{\rm end}^{\phantom1}&=&-1.258\pm0.034\,.
\end{eqnarray}

\noindent Also combining the set of galaxies of all the 10 runs in series~G, 
we obtained the best fit Schechter exponent 
for the surviving galaxy distribution at $z=0$ as $\alpha = -1.25$. 
Analogous to our approach for the $\beta$-model in \S\ref{sec-seriesC}, 
we note that $|\alpha|$ decreases by $\sim 0.03$ between $z=1$ and $z=0$. 
So we performed a new series of simulations, 
series~H, using $\alpha_{\rm start}=-1.31$, 
expecting that this will evolve to $\alpha \sim -1.28$ at $z=0$. 
This series includes galaxy harassment but no cD galaxy. 
The results for series~H are shown in Table~\ref{seriesh}. 

Series~H continues the trends of series~G pertaining to a NFW profile. 
There are a large number of halo tides that dominate the mass fraction, 
and result in a high value of $f_{\rm ICS}$. 
The combined numbers of tidal disruptions (by galaxy and halo) are 
comparable to or exceed the numbers of mergers. 
The mean $f_{\rm ICS}$ for series~H is
\begin{equation}
f_{\rm ICS}^{\phantom1}=0.471\pm0.129\,.
\end{equation}

In this series~H, we obtained the average values of $\alpha$ as 
\begin{eqnarray}
\alpha_{\rm start}^{\phantom1}&=&-1.310\pm0.020\,,\\
\alpha_{\rm end}^{\phantom1}&=&-1.288\pm0.036\,.
\end{eqnarray}

We then performed a series of simulations by putting a cD galaxy 
at the center of the NFW cluster halo, and called it series~I. 
The results are shown in Table~\ref{seriesi}. 
Here, the numbers of tides by the cluster halo and by other galaxies are comparable; 
when added the total occurrence of tides 
compares to or exceeds that of mergers. 
Comparing series~H and series~I (NFW-type clusters respectively without and with a cD galaxy,) 
there are more accretions when a cD galaxy is introduced (similar to series~D and F). 
The trend seen before with the Perseus-like clusters (between series~E and F), 
that the number of tidal disruptions by the background halo reduces 
in the presence of a cD galaxy, is almost absent in the NFW-type clusters.
Galaxies approaching the cluster center get destroyed by the tidal field of
the halo before the cD galaxy can have any effect. 

The galactic mass fractions dispersed into the ICM 
are neither too high, nor too low. 
We suspect this is the combined effect of putting a cD galaxy in a NFW type cluster. 
There is a tendency of the ICS mass fraction to be high in a NFW model cluster, 
and a cD galaxy tends to reduce the mass fraction imparted to the ICM. 
These two opposing trends cause the $f_{\rm ICS}$ values to be 
moderate in series~I. Here the mean $f_{\rm ICS}$ is 
\begin{equation} 
f_{\rm ICS}^{\phantom1}=0.302\pm0.044\,. 
\end{equation} 

\section{DEPENDENCE ON PARAMETERS OF CLUSTER HALO DENSITY PROFILE}
\label{sec-dependence} 

\subsection{$\beta$-Model} 
\label{sec-depend-beta} 

We investigated the dependence of the galaxy outcomes and 
the ICS mass fraction on the parameters governing the cluster halo density profile. 
For the $\beta$-model halo density (see \S\ref{sec-haloRho}), 
we found that the typical values of the relevant parameters cover the range 
$\beta = 0.3 - 1$, the core radius $r_c = 3 - 600$ kpc, 
the core number density $n_0 = (0.1-100) \times 10^{-3}$ cm$^{-3}$, 
corresponding to a physical density 
$\rho_0 = (0.017-17) \times 10^{-26}$ g cm$^{-3}$. 
We obtained these values from several observational studies, 
\citet{lea73, abramopoulos83, jones84, waxman95, makinoetal98, 
girardi98, ettori00, xue00, biviano03, piffarettietal06, maughan07}. 

We performed a series of simulations, as shown in Table~\ref{betaparams}, 
doing 5 runs in each series, for a total of 60 simulations. 
These were done using a Schechter mass distribution exponent 
$\alpha = -1.36$ (see \S\ref{sec-seriesC}), 
and including galaxy harassment, but not including a cD galaxy. 

\begin{deluxetable}{cccc}
\tabletypesize{\scriptsize}
\tablecaption{Series for Parameter Variations of $\beta$-Model Halo Density Profile}
\tablewidth{0pt}
\tablehead{
\colhead{Series} & \colhead{$\beta$} & 
\colhead{$\rho_0$ [g cm$^{-3}$]} & \colhead{$r_c$ [kpc]} 
}
\startdata
Bb1 & 0.3 & $1.0 \times 10^{-26}$ & 50 \\ 
Bb2 & 0.4 & $1.0 \times 10^{-26}$ & 50 \\
Bb3 & 0.5 & $1.0 \times 10^{-26}$ & 50 \\
Bb4 & 0.6 & $1.0 \times 10^{-26}$ & 50 \\
Bb5 & 0.8 & $1.0 \times 10^{-26}$ & 50 \\
Bb6 & 1.0 & $1.0 \times 10^{-26}$ & 50 \\ 
\hline 
Br1 & 0.5 & $1.0 \times 10^{-26}$ & 10 \\
Br2 & 0.5 & $1.0 \times 10^{-26}$ & 100 \\
Br3 & 0.5 & $1.0 \times 10^{-26}$ & 200 \\
Br4 & 0.5 & $1.0 \times 10^{-26}$ & 300 \\
Br5 & 0.5 & $1.0 \times 10^{-26}$ & 400 \\
Br6 & 0.5 & $1.0 \times 10^{-26}$ & 500 \\ 
\enddata
\label{betaparams}
\end{deluxetable} 

\begin{figure}
\begin{center}
\includegraphics[width=3.3in]{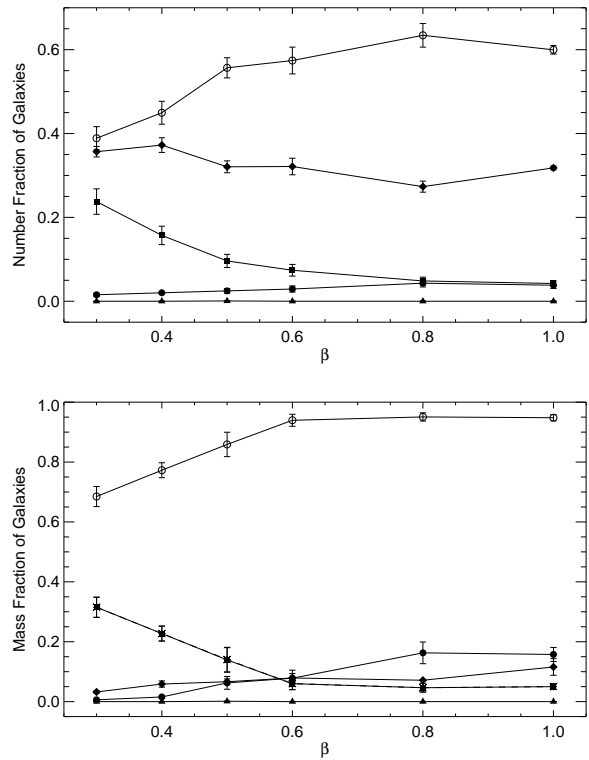} 
\caption{Number fractions (upper panel) and mass fractions (lower panel) 
of galaxies having different outcomes as a function of parameter $\beta$ 
of $\beta$-model halo density profile (see \S\ref{sec-depend-beta}). 
The symbols show the mean galaxy fractions and 
the error bars are the errors on the mean values. 
The common plotting symbols for both panels are: 
{\it diamond} - merger, 
{\it square} - tidal destruction by a massive galaxy with fragments going to ICM, 
{\it filled circle} - accretion after galaxy tidal disruption, 
{\it triangle} - tidal destruction by the halo, 
{\it open circle} - survival up to the present. 
In addition, the lower panel shows the galactic mass fraction that 
ends up in the ICM, 
$f_{\rm ICS}$, with {\it asterisk} plotting symbol, and joined 
by a {\it dashed} line. On this panel, the asterisks and the dashed line
are indistinguishable from the squares and the line connecting them.} 
\label{res_vs_beta}
\end{center}
\end{figure}

In series Bb1 -- Bb6 (Table~\ref{betaparams}), 
we explored values of $\beta$ between 0.3 -- 1.0, 
with fixed $r_c = 50$ kpc, and $\rho_0 = 1.0 \times 10^{-26}$ g cm$^{-3}$ 
(which corresponds to a particle number 
density $n_0 = 6.0 \times 10^{-3}$ cm$^{-3}$). 
We found the arithmetic means and the uncertainties on the arithmetic means ($=\sigma/\sqrt{5}$) 
of the galaxy fraction results from the 5 runs in each series. 
The resulting average fractions of each outcome 
are plotted in Figure~\ref{res_vs_beta} as a function of $\beta$, 
with the galaxy number fractions in the upper panel, 
and the mass fractions in the lower panel. 
Occurrences of ejection out of the cluster (\S\ref{sec-eject}) are not shown, 
since they are either zero or negligibly small compared to the other outcomes. 
As a note, for a certain value of $\beta$, 
the number fractions of the different galaxy outcomes will add up to 1. 
But the added mass fraction of all the outcomes will be $> 1$, 
because there is a double counting of the mass of galaxies merging and accreting. 
The mass fraction of galaxies surviving, the mass imparted to the ICM by tidal disruptions 
(due to other galaxies and cluster halo), and that ejected, will add up to 1. 

{From} Figure~\ref{res_vs_beta} we see that an increase in $\beta$ 
causes a decline of galaxy interactions, 
leading to survival of more and more galaxies up to $z=0$. 
The numbers of mergers dominate over galaxy tides by a factor of 2 or more. 
But only the low-mass galaxies merge, making the merged mass fraction 
significantly smaller than the tidal destruction mass fraction when $\beta < 0.6$. 
With increase of $\beta$ from 0.3 to 0.6, 
destruction by galaxy tides 
(with the fragments imparted to the ICM) decline dramatically, 
mergers have a slight decrease in number but increase in mass 
(implying that more massive galaxies are merging), 
accretions increase but always remaining below mergers and galaxy tides. 
In between $\beta = 0.6-1.0$ there is not much change in the resulting galaxy fractions, 
the trends continue slowly as $\beta$ reaches 1. 
A noteworthy result is that at $\beta > 0.6$ galaxy accretion dominates 
the mass fraction compared to the other outcomes. 
Tidal disruption by the halo and ejection out of the halo are either zero, 
or negligibly small in number and mass, for all $\beta$. 

The mass fraction $f_{\rm ICS}$ of galaxies imparted to the ICM by tidal disruptions 
is shown in the lower panel of Figure~\ref{res_vs_beta} 
with {\it asterisk} symbols joined by a {\it dashed} line. 
This curve is not distinguishable since, for this set of simulations, 
the ICS mass comes only from the dwarf galaxies destroyed by the tides of more massive galaxies. 
So $f_{\rm ICS}$ coincides with the {\it solid} line showing the galaxy tides fraction, 
with the {\it asterisk} symbols coinciding with the {\it squares}. 
As $\beta$ rises between $0.3-0.6$, 
the ICS mass fraction decreases from $\sim 0.3$ to $\sim 0.1$. 
After that $f_{\rm ICS}$ is essentially constant at $\sim 0.1$ when $\beta > 0.6$. 

\begin{figure}
\begin{center}
\includegraphics[width=3.3in]{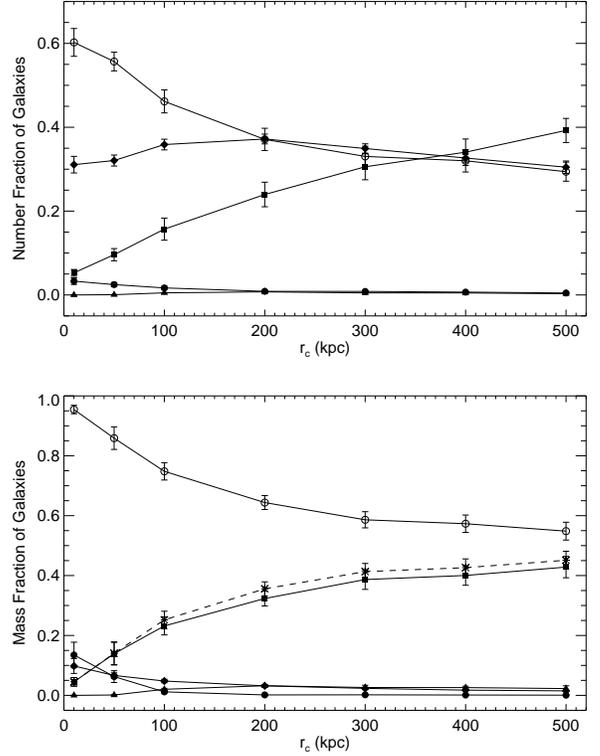}
\caption{Number fractions (upper panel) and mass fractions (lower panel) 
of galaxies having different outcomes, 
and mass fraction imparted to the ICM, $f_{\rm ICS}$ (lower panel), 
as a function of the core radius, $r_c$, of $\beta$-model halo density profile. 
The plotting symbols are same as in Figure~\ref{res_vs_beta}. 
See \S\ref{sec-depend-beta} for discussions.} 
\label{res_vs_Rcore}
\end{center}
\end{figure}

Next, we explore variations of the core radius 
$r_c$ in between 10 -- 500 kpc, in series Br1 -- Br6 (Table~\ref{betaparams}). 
Here $\beta = 0.5$, and $\rho_0 = 1.0 \times 10^{-26}$ g cm$^{-3}$ are kept fixed. 
We calculated and analyzed the results in a way analogous to that done for series Bb1 -- Bb6. 
The means and the errors on the means (from the 5 random runs in each series) 
of the resulting galaxy fractions with a specific outcome
are plotted in Figure~\ref{res_vs_Rcore} as a function of $r_c$. 

{From} Figure~\ref{res_vs_Rcore}, the following trends can be inferred 
about the dependence of the results on the core radius. 
As $r_c$ rises, a decreasing number of galaxies survive up to $z=0$, 
leading to an increase in the mass incorporated to the ICM. 
When $r_c > 50$ kpc, the mass fraction is increasingly dominated by galaxies 
getting destroyed by other galaxy tides with the fragments being dispersed into the ICM. 
Mergers outnumber galaxy tides when $r_c$ is small, 
become comparable to tides at $r_c \sim 350$ kpc, 
and at large-$r_c$ tides dominate. 
But only the low-mass galaxies merge (similar to series Bb1 -- Bb6), 
causing the merged mass fraction to be significantly smaller than that tidally destroyed. 
As $r_c$ increases, 
galaxy tides increase substantially by number and mass, 
mergers remain almost constant with a small reduction at large-$r_c$, 
and accretions (which have a small contribution) decline. 
There are a few cases of tidal destruction by the cluster halo 
when $r_c \geq 100$ kpc, which has a few $\%$ contribution by number and mass. 
The number of occurrences of ejection from the halo is always negligibly small in number and mass. 

The lower panel of Figure~\ref{res_vs_Rcore} shows $f_{\rm ICS}$, the 
mass fraction transferred to the ICM by destruction due to tides, 
as {\it asterisk} symbols joined by a {\it dashed} line. 
$f_{\rm ICS}$ is largely dominated by the galactic mass destroyed by 
the tides of more massive galaxies ({\it squares} joined by {\it solid} lines). 
There is a small contribution coming from tidal disruptions by the cluster halo, 
which gets added to the resultant $f_{\rm ICS}$. 
With increase of $r_c$ between $10 - 500$ kpc, 
the ICS mass fraction rises from $\sim 0.05$ to $\sim 0.45$. 

\subsection{NFW Model}
\label{sec-depend-NFW} 

We explored the dependence of the galaxy outcomes and 
the ICS mass fraction on the parameters governing the NFW model 
cluster halo density profile (see \S\ref{sec-haloRho}). 
Typical ranges of the profile parameters cover values of 
the scale radius $r_s = 100 - 500$ kpc, and the concentration parameter $c = 3 - 6$. 
We obtained these values from several observational works, 
\citet{carlbergetal97, vanderMarel00, arabadjis02, pratt02, biviano03, 
pointecouteau05, pratt05, maughan07}. 

Table~\ref{NFWparams} shows the series of simulations we performed, 
doing 5 random runs in each series, for a total of 35 simulations. 
For this set, we used a Schechter mass distribution exponent $\alpha = -1.31$ 
(see reasoning in \S\ref{sec-seriesGHI}), 
included galaxy harassment, but did not include a cD galaxy. 

\begin{deluxetable}{ccc}
\tabletypesize{\scriptsize}
\tablecaption{Series for Parameter Variations of NFW Model Halo Density Profile}
\tablewidth{0pt}
\tablehead{
\colhead{Series} & \colhead{c} & \colhead{$r_s$ [kpc]} 
}
\startdata
Nr1 & 4.5 & 10 \\ 
Nr2 & 4.5 & 50 \\
Nr3 & 4.5 & 100 \\
Nr4 & 4.5 & 200 \\ 
Nr5 & 4.5 & 300 \\ 
\hline 
Nc1 & 4 & 100 \\ 
Nc2 & 6 & 100 \\ 
\enddata
\label{NFWparams}
\end{deluxetable} 

\begin{figure}
\begin{center}
\includegraphics[width=3.3in]{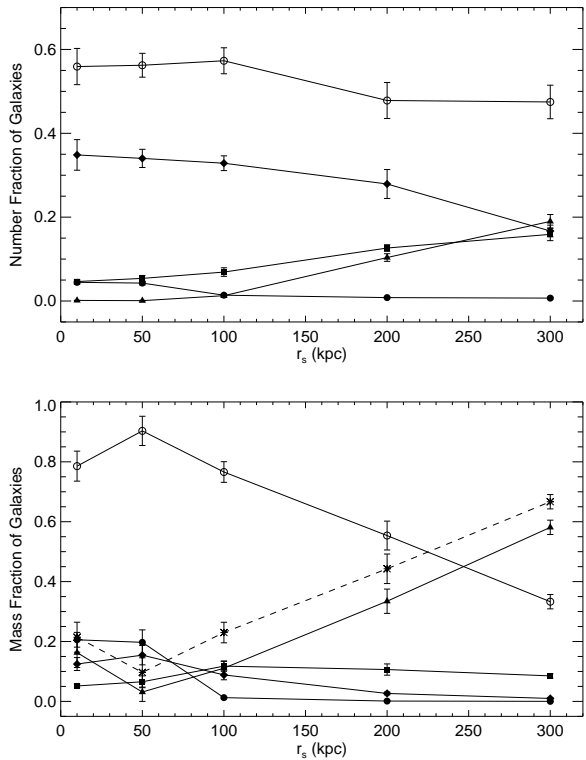}
\caption{Number fractions (upper panel) and mass fractions (lower panel) 
of galaxies having different outcomes, 
and mass fraction imparted to the ICM, $f_{\rm ICS}$ (lower panel), 
as a function of the scale radius, $r_s$, of NFW-model halo density profile. 
The plotting symbols are same as in Figure~\ref{res_vs_beta}. 
See \S\ref{sec-depend-NFW} for discussions.} 
\label{res_vs_Rscale}
\end{center}
\end{figure}

In series Nr1 -- Nr5 (Table~\ref{NFWparams}), 
we investigate five different values of the scale radius $r_s$ within 10 -- 300 kpc, 
keeping the concentration fixed at $c = 4.5$. 
We present the results in a manner analogous to that done for 
series Bb1 -- Bb6 and Br1 -- Br6 in \S\ref{sec-depend-beta}. 
Figure~\ref{res_vs_Rscale} shows the means and 
the errors on the means (from the 5 random runs in each series) 
of the resulting galaxy fractions with a specific outcome as a function of $r_s$. 

Some general trends are clear from Figure~\ref{res_vs_Rscale} 
about the dependence of the results on the scale radius. 
With increase of $r_s$, a smaller number of galaxies survive up to the present, 
causing greater mass transferred to the ICM. 
The most noteworthy feature of these NFW simulations is that 
a large galactic mass fraction is tidally destroyed by the cluster halo. 
In fact when $r_s \geq 100$ kpc, 
the mass fraction is increasingly dominated by galaxies 
getting destroyed by tidal field of the halo, 
with the fraction reaching as high as $\sim 0.6$ when $r_s = 300$ kpc. 
These disrupted galaxy fragments are unconditionally dispersed into the ICM, 
which increases the ICS mass fraction substantially. 

The number of mergers is higher than that of 
tides by galaxy and halo when $r_s$ is small; 
but these numbers of mergers, galaxy tides and halo tides 
become comparable at $r_s \sim 300$ kpc. 
At the same time, the galaxies merging have smaller masses 
(similar to the results in \S\ref{sec-depend-beta}), 
making the merged mass fraction comparable to that tidally destroyed by galaxies, 
both of which are hugely outweighed by halo tides. 
As $r_s$ increases, 
tidal disruption by the halo rise significantly, 
galaxy tides increase slightly in number but decrease in mass 
(implying lower mass galaxies are tidally destroyed by other galaxies), 
and mergers decline. 
Similar to the results in \S\ref{sec-depend-beta}, 
ejection out of the cluster is always negligibly small in number and mass. 

The galactic mass fraction imparted to the ICM by tidal destructions 
is shown in the lower panel of Figure~\ref{res_vs_Rscale}. 
As discussed in previous paragraphs, 
here $f_{\rm ICS}$ is largely dominated by the galactic mass destroyed by 
the tides of the cluster halo ({\it triangles} joined by {\it solid} lines). 
Tidal disruptions by other galaxies have a small contribution to $f_{\rm ICS}$. 
As $r_s$ rises between $10 - 300$ kpc, 
the ICS mass fraction grows from $\sim 0.1$ to $\sim 0.65$. 

Finally, we performed 2 series of simulations, Nc1 and Nc2 (Table~\ref{NFWparams}), 
varying the concentration parameter to $c = 4$ and $6$, 
with a fixed value of the scale radius $r_s = 100$ kpc. 
We found that the mergers continue to outnumber the tides. 
The average values of the ICS mass fractions are, 
$f_{\rm ICS} = 0.196$ with $c=4$, and 
$f_{\rm ICS} = 0.278$ with $c=6$.

\section{DISCUSSION}
\label{sec-discussion} 

\subsection{Mergers vs. Tides} 

To quantify the relative importance of destruction by tides and by
mergers, we calculated, for each run, the following fractional numbers:
\begin{eqnarray}
f_{\rm destroyed}^{\rm mergers}&=&{N_{\rm merge}+N_{\rm accr}
\over N_{\rm destroyed}}\,,\\
f_{\rm destroyed}^{\rm tides}&=&{N_{\rm tides}^{\rm gal}+N_{\rm tides}^{\rm halo}
\over N_{\rm destroyed}}\,,
\end{eqnarray}

\begin{figure}
\begin{center}
\vskip-2cm
\includegraphics[width=4in]{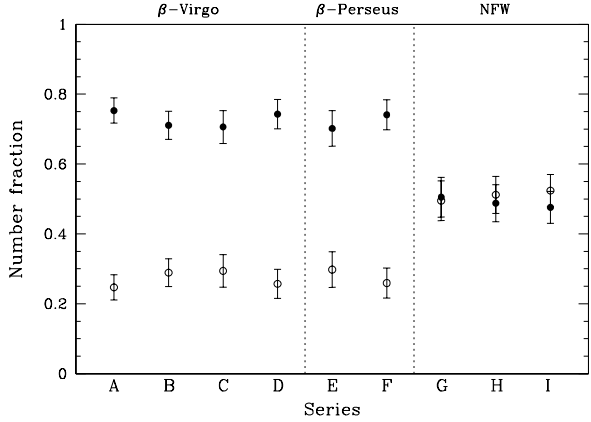}
\vskip-2.2cm
\caption{Fractional number of galaxies destroyed by mergers, 
$f_{\rm destroyed}^{\rm mergers}$ (filled circles), 
and by tides, $f_{\rm destroyed}^{\rm tides}$ (open circles), 
averaged over all runs within each series of Table~\ref{series}. 
Error bars show the standard deviation.}
\label{ratio_avg}
\end{center}
\end{figure}

\noindent where 
$N_{\rm destroyed}=N_{\rm merge}+N_{\rm tides}^{\rm gal}+
N_{\rm accr}+N_{\rm tides}^{\rm halo}$. 
We then averaged the fractions over all the runs in each series of the simulations. 
The results for the set of series in Table~\ref{series} 
are shown in Figure~\ref{ratio_avg}. 
The destruction by mergers clearly dominates over 
destruction by tides for the $\beta$ model, 
while they are of comparable importance for the NFW model. 

\begin{figure}
\begin{center}
\vskip-1.5cm
\includegraphics[width=3.6in]{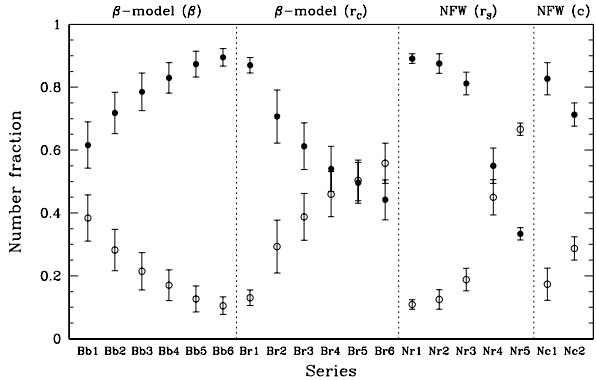}
\vskip-1.8cm
\caption{Fractional number of galaxies destroyed by mergers, 
$f_{\rm destroyed}^{\rm mergers}$ (filled circles), 
and by tides, $f_{\rm destroyed}^{\rm tides}$ (open circles), 
averaged over all runs within each series of 
Tables~\ref{betaparams} and \ref{NFWparams}. 
Error bars show the standard deviation.}
\label{ratio_avg2} 
\end{center} 
\end{figure}

Figure~\ref{ratio_avg2} gives the destruction fraction results for 
the set of series in Tables~\ref{betaparams} and \ref{NFWparams}. 
For greater values of the index $\beta$ of the $\beta$-model density 
(series Bb1--Bb6, Table~\ref{betaparams}), mergers rise and tides decline, 
causing mergers to increasingly dominate over tides. 
The trend is opposite for increasing core radius $r_c$ of the 
$\beta$-model (series Br1--Br6, Table~\ref{betaparams}), and 
for increasing scale radius $r_s$ of the NFW model 
(series Nr1--Nr5, Table~\ref{NFWparams}); 
here mergers reduce and tides grow, 
such that finally (at $r_c \geq 400$ kpc, and $r_s \geq 300$ kpc) 
tides outnumber mergers. 
Destruction by mergers well dominates that by tides 
in series Nc1 and Nc2 (Table~\ref{NFWparams}). 

\subsection{Intracluster Stars} 

Several mechanisms can contribute to removing stars 
from individual galaxies and putting them in the intracluster space. 
The efficiency and relative importance of these processes is expected to vary 
according to the location inside a cluster and during its evolution. 
If ram-pressure stripping or harassment are dominant mechanisms to 
produce IC stars, 
then the ICL fraction should increase with the mass of a cluster. 
On the other hand, if galaxy-galaxy merging is the dominant mechanism, 
and most of the ICL formed early on in cluster collapse, 
then the ICL fraction should be independent of present cluster mass. 

The ICL fraction depends on the merger history of the cluster, 
presence of a cD galaxy, and the morphology of the cluster galaxies. 
Observations \citep[e.g.,][]{krick07} show that 
if a cD galaxy is present then the ICL profile is centrally concentrated, 
implying that the ICL formed by galaxy interactions at the center, 
or formed earlier in protoclusters and later combined at the center. 
The ICL fraction should evolve with redshift, 
as the number of galaxy interactions increase with time. 
Cosmological simulations indicate that the ICL fraction does 
increase with time as clusters evolve \citep{willman04, rudick06}. 

The different formation mechanisms impart some 
distinct perceptible trends in the ICL. 
If most of the IC stars originate in initial cluster collapse, 
their distribution and kinematics should closely follow that of the 
galaxies in the cluster. 
If the ICL builds up slowly with time due to processes like 
``galaxy harassment,'' ``tidal stripping,'' etc., 
then a fraction of IC stars should be located in long streams along the 
orbits of parent galaxies. 
Cosmological simulations with very high resolution is required 
to address these issues numerically. 

\begin{deluxetable*}{ccccc}
\tabletypesize{\scriptsize}
\tablecaption{Observed values of the ICL mass as a fraction of the total cluster mass} 
\tablewidth{0pt}
\tablehead{
\colhead{Index} & \colhead{Cluster} & \colhead{$f_{\rm ICS}$ ($\%$)} & 
\colhead{$\Delta f_{\rm ICS}$ ($\%$)} & \colhead{Reference}
}
\startdata
1. & Coma & 50 & & \citet{bernstein95} \\ 
2. & Abell 1689 & 30 & & \citet{tyson95} \\ 
3. & Abell 1651 & $< 5$ & & \citet{gonzalez00} \\ 
4. & M96 (Leo) Group & $< 1.6$ & & \citet{castro-rodriguez03} \\ 
5. & HCG 90 & 45 & 5 & \citet{white03} \\ 
6. & Virgo & 15.8 & 8 & \citet{feldmeier2004b} \\ 
7. & A801   & 16 & 4.7 & \citet{feldmeier2004a} \\ 
8. & A1234 & 17 & 4.4 & \citet{feldmeier2004a} \\ 
9. & A1553 & 21 & 16 & \citet{feldmeier2004a} \\ 
10. & A1914 & 28 & 16 & \citet{feldmeier2004a} \\ 
11. & 93 clusters & 50 & 10 & \citet{lin04} \\ 
12. & 683 clusters & 10.9 & 5.0 & \citet{zibetti05} \\ 
13. & A4059  & 22 & 12 & \citet{krick07} \tablenotemark{a} \\ 
14. & A3880  & 14 & 6 & \citet{krick07} \\ 
15. & A2734  & 19 & 6 & \citet{krick07} \\ 
16. & A2556  & 6 & 5 & \citet{krick07} \\
17. & A4010  & 21 & 8 & \citet{krick07} \\
18. & A3888  & 13 & 5 & \citet{krick07} \\
19. & A3984  & 10 & 6 & \citet{krick07} \\
20. & A141   & 10 & 4 & \citet{krick07} \\
21. & AC 114 & 11 & 2 & \citet{krick07} \\
22. & AC 118 & 14 & 5 & \citet{krick07} \\ 
\enddata
\label{table_fICS} 
\tablenotetext{a}{\citet{krick07} measured these $f_{\rm ICS}$ values in the $r$ band.} 
\end{deluxetable*} 

\begin{figure}
\begin{center}
\includegraphics[width=3.5in]{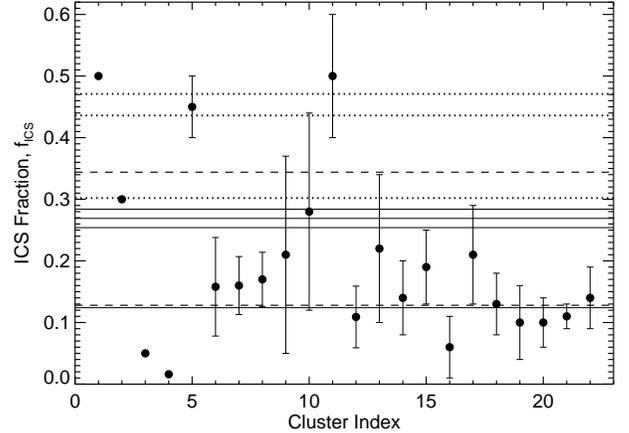}
\caption{ 
Fraction $f_{\rm ICS}^{\strut}$ of intracluster stars. 
The horizontal lines show the average $f_{\rm ICS}$ values 
in our simulations, 
from the runs of the series in Table~\ref{series} (\S\ref{sec-results}), with 
{\it solid line}: Virgo-like cluster (series A--D), 
{\it dashed line}: Perseus-like cluster (series E-F), 
{\it dotted line}: NFW model cluster (series G--I). 
The symbols and error bars show actual measurements, 
as tabulated in Table~\ref{table_fICS}. 
} 
\label{fics}
\end{center}
\end{figure}

There have been several observational measurements of the 
light (or, mass) fraction contained in the ICS with respect to 
the total light in a cluster. 
We collected some values of the ICS fraction from the literature, 
and list them in Table~\ref{table_fICS}. 
In Figure~\ref{fics} we show the ICS mass fraction 
we obtained in our simulations, 
plotted as horizontal lines showing the average $f_{\rm ICS}$ 
from the runs in Series A--I (Table~\ref{series}, \S\ref{sec-results}). 
For comparison, the observed $f_{\rm ICS}$ values 
(from Table~\ref{table_fICS}) are shown by the symbols and error bars. 
We can clearly see that the ICS mass fraction in clusters from observations 
fall well within our simulation predictions. 
A few clusters have too small $f_{\rm ICS}$, 
which are probably galaxy groups and low-mass clusters. 

Our simulation results indicate that 
the tidal destruction of dwarf galaxies (by other galaxies and by the cluster halo)
in clusters can sufficiently explain the observed fraction of intracluster light. 
Supporting our conclusions, observational studies do find a
significant ICL component arising from dwarf galaxies in clusters. 
{From} spectroscopic and photometric analysis, \citet{cote01} found that 
the  metal poor globular cluster system associated with M87 
(the cD galaxy in Virgo) 
is not formed in situ, but is stripped from the dwarf galaxies in Virgo. 
\citet{durrell02} observed that most of the IC red-giant branch stars in Virgo 
are moderately metal rich suggesting that they are 
tidally stripped from intermediate-luminosity galaxies, 
though stripping from DGs is not ruled out. 
\citet{mcnamara94} observed trails of of IC stars and gas from the 
tidal disruption of a DG by a companion giant elliptical galaxy in Virgo cluster. 

Results from our simulation runs (in \S\ref{sec-results} and \S\ref{sec-dependence}) 
indicate that, for each cluster halo density profile, namely, 
$\beta$, and NFW  models, 
$f_{\rm ICS}$ increases with the mass of the cluster halo. 
This is consistent with studies finding that more massive clusters have 
a larger fraction of ICL than the less massive ones \citep{lin04, murante04}. 

\subsection{Limitations of the Method}
\label{sec-limitations} 

The strengths and weaknesses of the methodology used in this work both
reside in our somehow original approach of using one single particle
to represent each galaxy, combined with a subgrid treatment of galaxy
mergers, tidal disruption, and galaxy harassment.

On the positive side, this approach has enabled us to perform a very
large number of simulations (\numberofrunstotal total), 
covering a fairly large parameter space,
while obtaining statistically significant results. In particular, the
dispersion of the results seen in Tables~\ref{seriesa}--\ref{seriesi}
justifies a posteriori the number of simulations we
have performed. Doing this many simulations
without resorting to subgrid physics would have been computationally
prohibitive. In addition, this approach greatly facilitates the
analysis and interpretation of the results. In a simulation in which
a galaxy would be represented by a large numbers of particles, a dwarf 
galaxy could be partly destroyed by tides, with some fragments being
dispersed in the ICM, other fragments accreting onto the larger
galaxy, with enough of the galaxy surviving that it can still be
counted as a galaxy. It would then become more difficult to
quantify objectively the relative importance of mergers and tidal disruptions.

In implementing the subgrid physics, we have attempted to make the 
most reasonable choices possible. One free parameter is the geometric factor
entering in equation~(\ref{internal}), but for reasonable density 
profiles, the values of that factor do not appear to vary much.
The assumption that a galaxy is considered ``tidally disrupted'' if 50\%
of its mass becomes unbound is also the most reasonable one we could
make. 

Our technique for generating the initial conditions is based
on four key assumptions: 
(1) the galaxy distributions are isotropic, 
(2) the galaxy number density profile $\nu(r)$ follows 
the density profile $\rho_{\rm halo}(r)$ of the background cluster halo, 
(3) the mass is segregated in the cluster, with the most massive 
galaxies being located in the center, and (4) the cluster is in 
equilibrium. So even though our prescription for generating the initial 
conditions contains many tunable parameters, we believe that the
underlying approach is sound.

On the negative side, two particular aspects of the methodology
can be considered weak. First, the treatment of galaxy harassment is
highly speculative. We have assumed that some amount of orbital
kinetic energy $\Delta E$ is dissipated into internal energy during
an encounter between two galaxies, that this amount is related to the
initial internal energies of the galaxies, and that the energy
dissipated is distributed equally between the two galaxies. The
dissipation of energy and its consequences during a real galactic encounter
are certainly much more complex, so the subgrid model could potentially
be refined. But this would require an intensive study of high-speed 
encounters, with detailed numerical simulations, and such study remains
to be done.

Another important limitation of our approach is that it deals with
isolated clusters in equilibrium. In the real universe, clusters
constantly experience mergers and accretion. We justify our approach
by the fact that most clusters will experience, at some epoch, a
{\it major merger}, during which most of the final mass of the cluster
is assembled. {From} that point, if we can neglect the addition of
mass by minor mergers and accretion, the cluster can be treated
as isolated.
Of course, such scenario cannot describe all clusters. In a 
forthcoming paper \citep{bmb08}, we will present a study
of cluster formation and evolution inside a cosmological volume
containing many clusters. This will be achieved by implementing
the subgrid approach described in this paper into a cosmological
N-body algorithm. 

\section{SUMMARY AND CONCLUSION} 
\label{sec-conclusion} 

We have designed a simple model for the evolution of galaxies 
in an isolated cluster, in order to compare the destruction of dwarf galaxies 
by mergers vs. tidal disruptions, and to predict the contribution 
of dwarf galaxies to the origin of intracluster stars. 
Our algorithm combines a direct N-body computation of gravitational interactions, 
along with a subgrid treatment of the other physical processes 
(merger, tidal disruption, accretion etc.) of the galaxies. 
Using this algorithm, we have performed a total of 
\numberofrunstotal numerical simulations of galaxy clusters, 
examining the fate of dwarf galaxies. 
Our results and conclusions are as follows. 

(1) The destruction of dwarf galaxies by mergers dominates over 
destruction by tides, in most of our simulation runs with all the models 
($\beta$-Virgo, $\beta$-Perseus, NFW) of cluster halo density. 
The two destruction mechanisms become comparable, and tides 
outnumber mergers, for the NFW and the $\beta$- models when their 
scale/core radius approach and exceed $\sim$ 200-300 kpc. 

(2) The destruction of dwarf galaxies by the tidal field of other galaxies 
and by the cluster halo imparts a significant amount of galactic mass into the ICM. 
This is sufficient to account for the observed fraction of 
intracluster light in galaxy clusters. 
In our simulations, the ICS mass fraction, $f_{\rm ICS}$, 
has a range $4.5 \% - 66.7 \%$ 
(with the majority of values being in between $10 \% - 45\%$), 
for the different parameter sets of the $\beta$ 
and NFW models of cluster halo density we considered. 
We see a clear trend of increase of $f_{\rm ICS}$ with the mass of the cluster halo. 
All these are well consistent with observations and other numerical studies. 

(3) In the NFW model simulated clusters, 
there are a large number of tidal disruptions of galaxies 
caused by the gravitational potential of the cluster halo, 
and this component dominates the mass fraction. 
We note that it has been our assumption that the cluster halo is stationary, 
and does not evolve in response to the forces exerted on it 
by the galaxies (\S\ref{sec-PPalgo}). 
Such an assumption is probably a poor one with the NFW model clusters. 
We point out that this could imply a possible solution 
to the cusp crisis of cluster dark matter halos. 
The central cuspy region of the cluster dark matter halo could have 
inelastic encounters with the member galaxies, 
which could inject energy into the halo and erase the cusp. 


(4) In our simulations, the presence of a cD galaxy increases occurrences of accretion,
decreases tidal disruptions by the cluster halo, and reduces the ICS mass fraction. 
This is opposite to the trend seen from observations that 
$f_{\rm ICS}$ is higher in the presence of a cD. 


\acknowledgments 

This work benefited from stimulating discussions with L. Edwards and C. Robert. 
We thank John Kormendy for useful correspondence. 
All calculations were performed at the Laboratoire 
d'astrophysique num\'erique, Universit\'e Laval. 
We thank the Canada Research Chair program and NSERC for support.

\appendix

\section{The Internal Energy of Galaxies}

Since we represent galaxies as individual particles, we cannot
directly compute their internal energy. We therefore need an
estimate that can then be used in equation~(\ref{internal}). We write the
potential energy of a galaxy of mass $M$ and radius $R$ as
\begin{equation}
W=-{\zeta GM^2\over R}\,,
\end{equation}

\noindent where $\zeta$ is the {\it geometric factor}, which depends
on the shape and density distribution of the galaxy. For a uniform-density
sphere, $\zeta=3/5$. In our simplified model, we treat galaxies as
spheres, but we should certainly not assume a uniform density. Instead,
any galaxy will be centrally concentrated. The value of $\zeta$ will
depend on the assumed density profile, but we do not expect
that dependence to be very strong if we stick with reasonable profiles.
So we consider the simplest case of an isothermal sphere with a cutoff
radius $R$. The density and mass inside $r$ are given by
\begin{eqnarray}
\rho(r)&=&{M\over4\pi Rr^2}\,,\\
m(r)&=&{Mr\over R}\,,
\end{eqnarray}

\noindent where $M\equiv m(R)$ is the total mass. The gravitational field
is given by
\begin{eqnarray}
{\bf g}=-\nabla\phi=-{Gm(r)\over r^2}\hat r=-{GM\over rR}\hat r\,.
\end{eqnarray}

\noindent We integrate this expression, with the boundary condition
$\phi(R)=-GM/R$, to get the gravitational potential,
\begin{equation}
\phi={GM\over R}\left(\ln{r\over R}-1\right)\,.
\end{equation}

\noindent The potential energy is given by
\begin{equation}
W={1\over2}\int\!\!\!\int\!\!\!\int\phi(r)\rho(r)d^3r=
{GM^2\over2R^2}\int_0^R\left(\ln{r\over R}-1\right)dr=-{GM^2\over R}\,.
\end{equation}

\noindent Hence, $\zeta=1$ for an isothermal sphere. Interestingly, this
is not much different from the value of 3/5 for a uniform sphere. This
supports our claim that the sensitivity of $\zeta$ on the density
profile is weak. For the kinetic energy, we assume that the galaxies
are virialized. Hence, $K=-W/2$, and therefore the internal energy is
given by
\begin{equation}
U=K+W=-{GM^2\over2R}\,.
\end{equation}

%


\clearpage

\begin{thebibliography}{} 

\bibitem[Abramopoulos \& Ku(1983)]{abramopoulos83} 
Abramopoulos, F., \& Ku, W. H. M. 1983, ApJ, 271, 446 

\bibitem[Aguerri et al.(2005)]{aguerri05} 
Aguerri, J. A. L. et al. 2005, AJ, 129, 2585 

\bibitem[Allen et al.(2002)]{allen02} 
Allen, S. W., Schmidt, R. W. \& Fabian, A. C. 2002, MNRAS, 334, L11 

\bibitem[Arabadjis, Bautz, \& Garmire(2002)]{arabadjis02} 
Arabadjis, J. S., Bautz, M. W., \& Garmire, G. P. 2002, ApJ, 572, 66 

\bibitem[Arnaboldi et al.(2003)]{arnaboldi03}
Arnaboldi, M. et al. 2003, AJ, 125, 514 

\bibitem[Arnaboldi(2004)]{arnaboldi04} 
Arnaboldi, M. 2004, IAUS, 217, 54 

\bibitem[Bernstein et al.(1995)]{bernstein95} 
Bernstein, G. M., Nichol, R. C., Tyson, J. A., Ulmer, M. P., \& Wittman, D.
1995, AJ, 110, 1507 

\bibitem[Biviano \& Girardi(2003)]{biviano03} 
Biviano, A., \& Girardi, M. 2003, ApJ, 585, 205 

\bibitem[Bothun et al.(1991)]{bothun91} 
Bothun, G. D. et al. 1991, ApJ, 376, 404 

\bibitem[Brainerd \& Specian(2003)]{brainerdetal03}
Brainerd, T. G., \& Specian, M. A. 2003, ApJ, 593, L7 

\bibitem[Brito et al.(2008)]{bmb08}
Brito, W., Martel, H., \& Barai, P. 2008, in preparation

\bibitem[Byrd \& Valtonen(1990)]{byrd90} 
Byrd, G., \& Valtonen, M. 1990, ApJ, 350, 89 

\bibitem[Carlberg et al.(1997)]{carlbergetal97}
Carlberg, R. G. et al. 1997, ApJ, 485, L13 

\bibitem[Carrasco et al.(2001)]{carrasco01} 
Carrasco, E. R. et al. 2001, AJ, 121, 148 

\bibitem[Castro-Rodriguez et al.(2003)]{castro-rodriguez03} 
Castro-Rodriguez, N., Aguerri, J. A. L., Arnaboldi, M., Gerhard, O., 
Freeman, K. C., Napolitano, N. R. \& Capaccioli, M. 2003, A\&A, 405, 803 

\bibitem[Cavaliere \& Fusco-Femiano(1976)]{cavaliereetal76}
Cavaliere, A., \& Fusco-Femiano, R. 1976, A\&A, 49, 137 

\bibitem[Cellone \& Buzzoni(2005)]{cellone05} 
Cellone, S. A., \& Buzzoni, A. 2005, MNRAS, 356, 41 

\bibitem[C\^ot\'e et al.(1997)]{cote97} 
C\^ot\'e, S., Freeman, K. C., Carignan, C., \& Quinn, P. J. 1997, AJ, 114, 1313 

\bibitem[C\^ot\'e, Carignan, \& Freeman(2000)]{cote00} 
C\^ot\'e, S., Carignan, C., \& Freeman, K. C. 2000, AJ, 120, 3027 

\bibitem[C\^ot\'e et al.(2001)]{cote01} 
C\^ot\'e, P. et al. 2001, ApJ, 559, 828 

\bibitem[De Propris et al.(2003)]{deproprisetal2003}
De Propris, R. et al. 2003, MNRAS, 342, 725 

\bibitem[Drinkwater et al.(2003)]{drinkwater03} 
Drinkwater, M. J. et al. 2003, Nature, 423, 519 

\bibitem[Durrell et al.(2002)]{durrell02} 
Durrell, P. R. et al. 2002, ApJ, 570, 119 

\bibitem[Ettori(2000)]{ettori00} 
Ettori, S. 2000, MNRAS, 318, 1041 

\bibitem[Ettori(2003)]{ettori03} 
Ettori, S. 2003, MNRAS, 344, L13 

\bibitem[Feldmeier et al.(2004a)]{feldmeier2004a} 
Feldmeier, J. J., Mihos, J. C., Morrison, H. L., Harding, P., Kaib, N.,
\&  Dubinski, J. 2004a, ApJ, 609, 617

\bibitem[Feldmeier et al.(2004b)]{feldmeier2004b} 
Feldmeier, J. J., Ciardullo, R., Jacoby, G. H., \& 
Durrell, P. R. 2004b, ApJ, 615, 196 

\bibitem[Ferguson \& Sandage(1991)]{ferguson91} 
Ferguson, H. C., \& Sandage, A. 1991, AJ, 101, 765 

\bibitem[Ferguson et al.(1998)]{ferguson98} 
Ferguson, H. C., Tanvir, N. R., \& von Hippel, T. 1998, Nature, 391, 461 

\bibitem[Gal-Yam et al.(2003)]{galyam03} 
Gal-Yam, A. et al. 2003, AJ, 125, 1087 

\bibitem[Gerhard et al.(2005)]{gerhard05} 
Gerhard, O. et al. 2005, ApJ, 621, L93 

\bibitem[Girardi et al.(1998)]{girardi98} 
Girardi, M., Giuricin, G., Mardirossian, F., 
Mezzetti, M., \& Boschin, W. 1998, ApJ, 505, 74 

\bibitem[Gnedin(2003)]{gnedin03} 
Gnedin, O. Y. 2003, ApJ, 589, 752 

\bibitem[Gonzalez et al.(2000)]{gonzalez00} 
Gonzalez, A. H., Zabludoff, A. I., Zaritsky, D. 
\& Dalcanton, J. J. 2000, ApJ, 536, 561 

\bibitem[Gonzalez et al.(2005)]{gonzalez05} 
Gonzalez, A. H., Zabludoff, A. I., \& Zaritsky, D. 2005, ApJ, 618, 195 

\bibitem[Gonzalez et al.(2007)]{gonzalez07} 
Gonzalez, A. H., Zaritsky, D., \& Zabludoff, A. I. 2007, preprint (arXiv0705.1726) 

\bibitem[Grebel(2001)]{grebel01} 
Grebel, E. K. 2001, ASPC, 239, 280 

\bibitem[Gregg \& West(1998)]{gregg98} 
Gregg, M. D., \& West, M. J. 1998,  Nature, 396, 549 


\bibitem[Hilker et al.(1999)]{hilker99} 
Hilker, M., Infante, L., \& Richtler, T. 1999, A\&AS, 138, 55

\bibitem[Hill \& Oegerle(1998)]{hill98} 
Hill, J. M., \& Oegerle, W. R. 1998, AJ, 116, 1529 

\bibitem[Impey et al.(1988)]{impey88} 
Impey, C. et al. 1988, ApJ, 330, 634 

\bibitem[Jones \& Forman(1984)]{jones84} 
Jones, C., \& Forman, W. 1984, ApJ, 276, 38 

\bibitem[Jordan et al.(2004)]{jordan04} 
Jordan, A., C\^ot\'e, P., West, M. J., Marzke, R. O., Minniti, D.,
\& Rejkuba, M. 2004, AJ, 127, 24 

\bibitem[Karachentseva et al.(1985)]{karachentseva85} 
Karachentseva, V. E., Karachentsev, I. D., \& Boerngen, F. 1985, A\&AS, 60, 213

\bibitem[Karachentsev et al.(2004)]{karachentsev04} 
Karachentsev, I. D., Karachentseva, V. E., Huchtmeier, W. K.,
\& Makarov, D. I. 2004, AJ, 127, 2031

\bibitem[Kauffmann, White, \& Guiderdoni(1993)]{kauffmann93} 
Kauffmann, G., White, S. D. M., \& Guiderdoni, B. 1993, MNRAS, 264, 201

\bibitem[King(1962)]{king62}
King, I. 1962, AJ, 67, 471

\bibitem[Krick et al.(2006)]{krick06} 
Krick, J. E., Bernstein, R. A., \& Pimbblet, K. A. 2006, AJ, 131, 168 

\bibitem[Krick \& Bernstein(2007)]{krick07} 
Krick, J. E., \& Bernstein, R. A. 2007, AJ, 134, 466

\bibitem[Lea et al.(1973)]{lea73} 
Lea, S. M., Silk, J., Kellogg, E., \& Murray, S. 1973, ApJ, 184, L105 

\bibitem[Lee et al.(2003)]{lee2003} 
Lee, H. et al. 2003, AJ, 125, 2975 

\bibitem[Lewis et al.(2002)]{lewis02}
Lewis, I. et al. 2002, MNRAS, 334, 673L 

\bibitem[Lin \& Mohr(2004)]{lin04} 
Lin, Y.-T., \& Mohr, J. J. 2004, ApJ, 617, 879 

\bibitem[Makino, Sasaki, \& Suto(1998)]{makinoetal98}
Makino, N., Sasaki, S., \& Suto, Y. 1998, ApJ, 497, 555

\bibitem[Martel(1991)]{martel91}
Martel, H. 1991, ApJ, 377, 7 

\bibitem[Mateo(1998)]{mateo98} 
Mateo, M. L. 1998, ARA\&A, 36, 435 

\bibitem[Maughan et al.(2007)]{maughan07}  
Maughan, B. J., Jones, C., Jones, L. R.,  \& Van Speybroeck, L. 
2007, ApJ, 659, 1125 

\bibitem[Mayer et al.(2001)]{mayer01} 
Mayer, L. et al. 2001, ApJ, 547, L123 

\bibitem[McNamara et al.(1994)]{mcnamara94} 
McNamara, B. R. et al. 1994, AJ, 108, 844 

\bibitem[Merritt(1984)]{merritt84} 
Merritt, D. 1984, ApJ, 276, 26 

\bibitem[Mieske et al.(2007)]{mieske07} 
Mieske, S., Hilker, M., Jordan, A., Infante, L. \& 
Kissler-Patig, M. 2007, preprint, arXiv:0706.2724 

\bibitem[Mihos(2004)]{mihos04} 
Mihos, J. C. 2004, in Clusters of Galaxies: Probes of Cosmological Structure 
and Galaxy Evolution, ed. J. S. Mulchaey, A. Dressler \& A. Oemler 
(Cambridge: Cambridge Univ. Press), 277 

\bibitem[Mihos et al.(2005)]{mihos05} 
Mihos, J. C., Harding, P., Feldmeier, J., \& Morrison, H. 2005, ApJ, 631, L41 

\bibitem[Miller(1983)]{miller83} 
Miller, G. E. 1983, ApJ, 268, 495 

\bibitem[Moore et al.(1996)]{moore96} 
Moore, B., Katz, N., Lake, G., Dressler, A., \& Oemler, A. 1996, Nature, 379, 613 

\bibitem[Mori \& Burkert(2000)]{mori00} 
Mori, M., \& Burkert, A. 2000, ApJ, 538, 559 

\bibitem[Murante et al.(2004)]{murante04} 
Murante, G. et al. 2004, ApJ, 607, L83 

\bibitem[Napolitano et al.(2003)]{napolitano03} 
Napolitano, N. R. et al. 2003, ApJ, 594, 172

\bibitem[Navarro, Frenk, \& White(1997)]{NFW97} 
Navarro, J. F., Frenk, C. S., \& White, S. D. M. 1997, ApJ, 490, 493 

\bibitem[Oegerle \& Hill(2001)]{oegerle01} 
Oegerle, W. R., \& Hill, J. M. 2001, AJ, 122, 2858 

\bibitem[Phillipps et al.(1998)]{phillipps98} 
Phillipps, S. et al. 1998, ApJ, 493, L59 

\bibitem[Piffaretti \& Kaastra(2006)]{piffarettietal06}
Piffaretti, R., \& Kaastra, J. S. 2006, A\&A, 453, 423 

\bibitem[Pointecouteau, Arnaud \& Pratt(2005)]{pointecouteau05} 
Pointecouteau, E., Arnaud, M. \& Pratt, G. W. 2005, A\&A, 435, 1 

\bibitem[Pratt \& Arnaud(2002)]{pratt02} 
Pratt, G. W. \& Arnaud, M. 2002, A\&A, 394, 375 

\bibitem[Pratt \& Arnaud(2005)]{pratt05} 
Pratt, G. W., \& Arnaud, M. 2005, A\&A, 429, 791 

\bibitem[Quintana \& Lawrie(1982)]{quintana82} 
Quintana, H., \& Lawrie, D. G. 1982, AJ, 87, 1 

\bibitem[Richstone(1976)]{richstone76} 
Richstone, D. O. 1976, ApJ, 204, 642 

\bibitem[Richstone \& Malumuth(1983)]{richstone83} 
Richstone, D. O., \& Malumuth, E. M. 1983, ApJ, 268, 30 

\bibitem[Rudick et al.(2006)]{rudick06} 
Rudick, C. S., Mihos, J. C., \& McBride, C. 2006, ApJ, 648, 936 

\bibitem[Ryan et al.(2007)]{ryanetal07} 
Ryan, R. E. Jr. et al. 2007, ApJ, in press (astro-ph/0703743)

\bibitem[Sandage et al.(1985)]{sandage85} 
Sandage, A., Binggeli, B., \& Tammann, G. A. 1985, AJ, 90, 1759 

\bibitem[Schechter(1976)]{schechter76}
Schechter, P. 1976, ApJ, 203, 297 

\bibitem[Schombert(1988)]{schombert88} 
Schombert, J. M. 1988, ApJ, 328, 475 

\bibitem[Seigar, Graham, \& Jerjen(2007)]{seigar07} 
Seigar, M. S., Graham, A. W., \& Jerjen, H. 2007, MNRAS, in press 

\bibitem[Sommer-Larsen et al.(2005)]{sommerlarsen05} 
Sommer-Larsen, J. et al. 2005, MNRAS, 357, 478 

\bibitem[Spergel et al.(2007)]{spergeletal07} 
Spergel, D. N. et al. 2007, ApJS, 170, 377 

\bibitem[Staveley-Smith, Davies, \& Kinman(1992)]{staveley92} 
Staveley-Smith, L., Davies, R. D., \& Kinman, T. D. 1992, MNRAS, 258, 334 

\bibitem[Thompson \& Gregory(1993)]{thompson93} 
Thompson, L. A., \& Gregory, S. A. 1993, AJ, 106, 2197 

\bibitem[Trentham \& Hodgkin(2002)]{trentham02a} 
Trentham, N., \& Hodgkin, S. 2002, MNRAS, 333, 423 

\bibitem[Trentham \& Tully(2002)]{trentham02b} 
Trentham, N., \& Tully, R. B. 2002, MNRAS, 335, 712 

\bibitem[Trentham, Tully, \& Mahdavi(2006)]{trentham06} 
Trentham, N., Tully, R. B., \& Mahdavi, A. 2006, MNRAS, 369, 1375 

\bibitem[Tyson \& Fischer(1995)]{tyson95} 
Tyson, J. A. \& Fischer, P. 1995, ApJ, 446, L55 

\bibitem[van der Marel et al.(2000)]{vanderMarel00} 
van der Marel, R. P., Magorrian, J., Carlberg, R. G.; Yee, H. K. C., 
Ellingson, E. 2000, AJ, 119, 2038 

\bibitem[Vilchez-Gomez(1999)]{vilchez99}
Vilchez-Gomez, R. 1999, ASPC, 170, 349 

\bibitem[Waxman \& Miralda-Escude(1995)]{waxman95} 
Waxman, E., \& Miralda-Escude, J. 1995, ApJ, 451, 451 

\bibitem[Weil et al.(1997)]{weil97} 
Weil, M. L. et al. 1997, ApJ, 490, 664 

\bibitem[West et al.(1995)]{west95} 
West, M. J. et al. 1995, ApJ, 453, L77 

\bibitem[White \& Frenk(1991)]{white91}
White, S. D. M., \& Frenk, C. S. 1991, ApJ, 379, 52 

\bibitem[White et al.(1993)]{white93} 
White, S. D. M., Navarro, J. F., Evrard, A. E., \& Frenk, C. S. 1993, 
Nature, 366, 429 

\bibitem[White et al.(2003)]{white03} 
White, P. M., Bothun, G., Guerrero, M. A., 
West, M. J. \& Barkhouse, W. A. 2003, ApJ, 585, 739 

\bibitem[Willman et al.(2004)]{willman04} 
Willman, B., Governato, F., Wadsley, J., \& Quinn, T. 2004, MNRAS, 355, 159 

\bibitem[Xue \& Wu(2000)]{xue00} 
Xue, Y.-J., \& Wu, X.-P. 2000, MNRAS, 318, 715 

\bibitem[Zibetti et al.(2005)]{zibetti05} 
Zibetti, S., White, S. D. M., Schneider, D. P., \& Brinkmann, J. 2005, 
MNRAS, 358, 949 

\bibitem[Zwicky(1951)]{zwicky51} 
Zwicky, F. 1951, PASP, 63, 61

\end{thebibliography}
\end{document}